\documentclass[11pt,letterpaper]{article}
\usepackage[margin=1in]{geometry}
\usepackage{amsmath,amssymb,amsthm}
\usepackage{booktabs}
\usepackage{graphicx}
\usepackage[title]{appendix}
\usepackage[authoryear,round]{natbib}
\usepackage{url}
\usepackage[hidelinks]{hyperref}
\usepackage{microtype}

\newcommand{\alttext}[1]{}
\newcommand{\runin}[1]{\par\medskip\noindent\textit{#1}\enspace}
\newcommand{\Eo}{\textsf{E0}}
\newcommand{\Ei}{\textsf{E1}}
\newcommand{\Em}{\textsf{E1m}}
\newcommand{\ms}{\,\text{ms}}

\newcommand{\sx}{\,\text{s}}

\theoremstyle{remark}
\newtheorem{remark}{Remark}
\theoremstyle{definition}
\newtheorem{definition}{Definition}

\title{Event definition and the branching ratio: Hawkes estimates of price-change clustering in the consolidated quote feed}
\author{Hyoeun Lee\thanks{Department of Statistics, University of Illinois Urbana-Champaign, Champaign, IL 61820, USA. E-mail: \href{mailto:hyoeun@illinois.edu}{hyoeun@illinois.edu}.}}
\date{October 2026}

\begin{document}
\maketitle

\begin{abstract}
\noindent Hawkes models of price changes in the consolidated US quote feed find a sub-millisecond kernel component and a branching ratio near criticality. We show that in large-tick stocks the first is the feed's construction: half of NBBO changes are locks or depleted quotes, resolved within a millisecond, which the model reads as excitation. Defining the event as a mid-price level persisting for one millisecond removes the component. Raw fits put three times the clean kernel mass within one second, which a simulation reproduces. Since the total branching ratio trades off against the baseline, we report kernel mass on a fixed window.

\medskip
\noindent\textit{Keywords:} Hawkes process; branching ratio; consolidated NBBO; locked markets; TAQ; event definition; refractory period.

\noindent\textit{JEL classification:} C51, C58, G10, G14.
\end{abstract}

Hawkes processes are the standard statistical model for the clustering of events in financial markets: one price change makes another more likely soon after. When the model is fitted to the price changes of US stocks recorded in the consolidated quote feed, the Trade and Quote (TAQ) data of the NYSE, it finds two things that have become part of the empirical picture. First, the kernel needs a component with memory below one millisecond, to which about 40\% of all events are attributed \citep{lee2023}. Second, the branching ratio, the expected number of events that each event triggers, is high. The branching ratio lies between zero and one for a stationary process, and a value near one means that most events are triggered by earlier events and few arrive from outside. The same study finds 0.6--0.7 for IBM and AAPL, in the range that \citet{filimonov2012} and \citet{hardiman2013} read, on index futures, as a market close to criticality. \citet{filimonov2015}, however, show that estimates in this range can arise from the calibration itself, so the reading is not settled.

Both findings depend on how the consolidated feed records a price move, which is where this paper starts. A US stock trades on more than a dozen exchanges. Each exchange sends its quotes to a central processor, which assembles the national best bid and offer (NBBO) that TAQ records, and the quotes reach the processor with delays that differ by venue. In a stock whose spread is pinned at the one-cent minimum, a price move usually requires two venues to update. Consider a trader who wants to bid at the old ask price; on the venue that holds the ask, such a bid would execute as a trade, so the trader posts it on another venue. The venue holding the ask then updates its quote a fraction of a millisecond later. Exchanges act on their direct feeds faster than the processor consolidates them, so in between the processor publishes a locked market: a bid equal to the ask. About half of all NBBO changes in such stocks are locks of this kind. They last a median $0.4$--$0.7\ms$, and in INTC 96\% of price moves pass through one. There is a further class of brief states, in which the spread widens for tens of microseconds and returns. When we match these states to the trade record, they turn out to be the moment after a trade takes out a quote. In either case the mid-price moves twice within a millisecond where the price level moved once, or it moves and returns where the level did not move at all. A Hawkes model has only one way to explain two events a millisecond apart: excitation. The brief states are therefore fitted as excitation.

In this paper, we proceed in four steps. First, we define the target process as something observable: the mid-price changes whose new level persists for at least a threshold $h$, with $h = 1\ms$, stamped when the level is first quoted. Raw NBBO changes, changes with locks merged, and changes with every state shorter than $h$ merged are then three constructions from the same data, and the question becomes how the estimates depend on the construction. Second, we fit a bivariate Hawkes model to each of the three series. The kernel is a sum of exponentials, the baseline is constant within each half hour of each day, and a refractory period equal to $h$ follows every event. The sample is eight liquid stocks over fourteen months, November 2016 to December 2017, most of them in the large-tick regime. We compare the three series on the kernel mass within one second and within ten seconds. We choose this summary because the data identify it and because it does not depend on the baseline specification within one second, and within ten seconds only at the finest baselines, whereas the total branching ratio does throughout. Third, we simulate price moves from the fitted Hawkes model, add the brief states measured in the data through a synthetic consolidated feed, and fit the model again under each event definition. Each construction can then be compared with the values that generated it. Fourth, we examine what the clean series shows about price-change clustering, and where the model still fails. In what follows, we ask whether the sub-millisecond component is a feature of the price or of the feed, what the raw branching ratio measures and which summary of a fit survives the construction, what the clean series shows about price-change clustering and how uncertain a month's estimate is, and where the model fitted to the clean series still fails.

We apply this approach to the eight stocks and make the following findings. First, the sub-millisecond component is a product of how the feed is built, fragmentation across venues stretched by the consolidation, and not a feature of the price. Raw-event fits put about three times the kernel mass within one second that clean-event fits do, most of the difference lies in the first millisecond, and in price units the raw model implies close to three times the post-event price continuation that the data show. The simulation reproduces this distortion from the measured brief states alone. A genuine sub-millisecond component is recovered by the fit from raw events only when the feed has no brief states, which in consolidated data it always does. Second, the raw branching ratio is contaminated in a predictable way. It obeys a simple counting relation, the share of raw events that resolve a brief state, about one half, plus the real excitation diluted by about one half, so raw data compress the differences between stocks and months. In addition, the total branching ratio, raw or clean, trades off against the baseline across resolutions on which AIC and BIC disagree, whereas the kernel mass within one second does not, and within ten seconds does not from 30-minute bins upward. We therefore take the kernel mass within one and ten seconds as the quantity to report, and the simulation shows that the clean construction recovers it for a four-exponential and for a power-law kernel. Third, on the clean series the fitted kernel places its mass between about 10 ms and 30 s, and excitation rises with activity, across months, within the day and across days, while the brief states become shorter when the market is busy. A day-block bootstrap shows that this day-to-day variation, not estimation noise, is the uncertainty of a month's estimate. Finally, the model fitted to the clean series is still rejected by calibrated tests, and we locate the misfit in three places: stronger excitation at the open, bursts shorter than five minutes, and more reversals than a linear symmetric kernel can produce. The magnitudes behind these findings are given in Sections~\ref{sec:simulation} and~\ref{sec:empirical}.

The rest of the paper is organized as follows. Section~\ref{sec:literature} reviews the literature. Section~\ref{sec:nbbo} describes what the consolidated NBBO records during a price move, defines the target process, and explains why the definition matters for a Hawkes model. Section~\ref{sec:model} presents the model and its estimation, and Section~\ref{sec:analysis} the quantities we report and why. Section~\ref{sec:simulation} contains the simulation study and Section~\ref{sec:empirical} the empirical study, which begins with the data and the tick regime of each stock-month (Section~\ref{sec:data}). Section~\ref{sec:conclusion} concludes. The appendices give the gated likelihood and the fully gated kernel mass.

\section{Literature Review}
\label{sec:literature}

This paper draws on three strands of literature: Hawkes models of price changes and the branching ratio, the market-structure literature on the consolidated feed, and the methods used to estimate and test the model.

Hawkes processes \citep{hawkes1971, hawkesoakes1974} entered the microstructure literature as a model of order flow and price changes, and \citet{bacry2015} review the applications. Our specification descends from two papers. \citet{bacry2013} model up and down moves of a price as two mutually exciting types. \citet{leeseo2017} apply a symmetric version to the mid-price of US stocks observed from one-second NYSE quotes, a resolution at which no sub-millisecond component can appear. It is well established that financial kernels decay slowly, over several timescales. \citet{hardiman2013} and \citet{bacry2016} find power-law kernels on futures. \citet{lallouache2016} and \citet{lee2023} represent the decay by a sum of exponentials. The latter fits the NBBO mid-price of IBM and AAPL from TAQ and finds a component with memory below one millisecond that carries about 40\% of events.

The branching ratio as a measure of market endogeneity is due to \citet{filimonov2012}, and its difficulties are the subject of a continuing debate. \citet{filimonov2015} show that several features of the data and of the calibration each produce spurious excitation and apparent criticality. These are the bundling of messages into packets, which leaves the true time of an event uncertain within the packet interval, outliers in the inter-event times, regime shifts in the background intensity, and a non-exhaustive search of the likelihood. \citet{omi2017} fit a time-dependent background rate, and \citet{wehrli2021} document how the estimates depend on the sampling scale, the time window and the asset. Note that the mechanism in \citet{filimonov2015} distorts the times of events that are already in the series, on a feed from a single exchange. In our setting, the consolidation of several exchanges creates events, in pairs attached to real moves. Our results also sharpen their point about regime shifts in one respect. In Section~\ref{sec:tradeoff}, we show that the total branching ratio trades off against the baseline even when the baseline changes by half hour and by day, among resolutions that the information criteria cannot distinguish.

The consolidated feed itself has been studied extensively in market microstructure, whereas its effect on point-process models remains unexamined. \citet{shkilko2008} document locked and crossed markets and their resolution. \citet{holden2014} and \citet{favreau2022} treat the completed NBBO table and the handling of non-positive spreads, and \citet{hasbrouck2018} measures the short-term volatility of quotes. \citet{ding2014}, \citet{bartlett2019} and \citet{tivnan2020} measure the dislocations between the consolidated feed and the exchanges' direct feeds. \citet{holden2023} and \citet{battalio2026} show that exchange-to-SIP latency distorts trade classification and introduces look-ahead bias into trade-and-quote matching. Our exchange-time reconstruction of the NBBO follows \citet{holden2023}. The regulator's view is set out in the SEC's 2026 proposing release on the trade-through and locked-market provisions of Regulation NMS \citep{sec2026}. Citing \citet{mackintosh2025} and \citet{battalio2026}, the release attributes most locks in SIP data to latency between data centers and the SIP. The 2024 amendments reducing the tick size for tick-constrained stocks \citep{sec2024} had not been implemented at the time of writing. \citet{dayri2015} define the large-tick regime from which our sample is drawn. Operationally, we call a stock-month large-tick when its spread is one tick at least 99\% of the time (Section~\ref{sec:data}).

Three Hawkes papers on best-quote data come closest to ours. \citet{leeseo2023} model best bid and ask moves of high-priced stocks with a spread-dependent Hawkes process. They treat locked and crossed NBBOs as network-delay artifacts and remove them. In their sample these states are 1--2\% of events, and their effect is not measured. In large-tick stocks they are about half of all events. \citet{lee2024} filters NBBO mid-prices on a 0.1-second grid to remove the ultra-high-frequency activity that his earlier fits had attributed to automated quoting. His purpose is to estimate daily volatility. In Section~\ref{sec:robust}, we compare his grid with our duration rule and show that the grid removes the clustering at 10--100 ms along with the brief states. \citet{leelee2026} model the four best-quote moves with the spread as a state variable, on data from a single exchange, where no locks occur. Their cross-side kernel captures a millisecond snap-back of the spread, which is the kind of movement that our definition merges. The two approaches are complementary. For the quotes themselves such mechanics are structure, while for the clustering of price formation they are contamination. These studies remove the brief states or filter them out, whereas what the construction of the consolidated feed does to a Hawkes fit, and which summary survives it, remain unexamined. This paper fills that gap.

We use three methodological strands without extending them. The first is estimation of Hawkes processes under noisy or missing event times, treated by \citet{deutsch2020}, \citet{trouleau2021}, \citet{cheysson2022} and \citet{bonnet2025}. The feed poses a different problem: it does not perturb event times or add independent noise, but creates events attached to real ones, each within a millisecond of a genuine move as its pair. Because of this dependence, we address the problem by defining the event. The alternatives would be to correct the estimator, or to model the observation mechanism jointly with a latent process, which would require a model of the pairing that Section~\ref{sec:simulation} supplies only as an experiment. The second strand is the theory and inference of the Hawkes model itself. Bootstrap inference and goodness-of-fit testing follow \citet{cavaliere2023} and \citet{kling2026}. The likelihood with the exponential recursion is that of \citet{ozaki1979}, the cluster representation we rely on in Section~\ref{sec:analysis} is that of \citet{hawkesoakes1974} and \citet{daley2003}, and stationarity of the process follows \citet{bremaud1996}. The notion of excitation we estimate is the Granger-type one of \citet{eichler2017}, not a causal one. The third strand is data cleaning. The rules of the realized-volatility literature are the nearest existing remedy for a problem of this kind. \citet{barndorffnielsen2009} delete crossed quotes and outliers and, for the rest, recommend retaining the quotes of a single exchange, which avoids consolidation altogether. A Hawkes model of the national best quote cannot take that route, and the standard rules keep locked quotes and sub-millisecond states, so the artifact survives them.

\section{Events in the Consolidated NBBO}
\label{sec:nbbo}

In this section, we describe what the consolidated NBBO records while a price moves, define the process we model, and explain why the definition matters for a Hawkes model.

\subsection{Brief States: Locks and Depletion}
\label{sec:briefstates}

A US stock trades on more than a dozen exchanges, each with its own order book. Regulation NMS ties them together through the national best bid and offer (NBBO), the highest bid and lowest ask across all exchanges. A Securities Information Processor (SIP) computes the NBBO from the quotes every exchange sends it and publishes it. In 2016--17 the exchanges sat in three New Jersey data centers, Mahwah, Carteret and Secaucus. The two SIPs sat in two of them: the SIP for NYSE-listed stocks in Mahwah and the SIP for Nasdaq-listed stocks in Carteret. Their quotes reached the SIP after delays of tens to hundreds of microseconds that differ by venue (see \citealp{ding2014, tivnan2020, holden2023}). Rule 610(d) requires each exchange to adopt rules designed to prevent the display of quotes that lock the market (a bid equal to the best ask elsewhere) or cross it. Exchanges enforce these rules against their own direct feeds, which they see sooner than the SIP consolidates them. Nonetheless, the consolidated NBBO is locked or crossed many times a day, for very short periods, and the regulator attributes most of these locks to latency between the data centers and the SIP \citep{sec2026, mackintosh2025, battalio2026}.

\begin{figure}[!t]
\centering
\includegraphics[width=\linewidth, alt={Two rows of three price-ladder diagrams. Each diagram shows the best bid and best ask on venues X and Y and the NBBO composed from them, before, during and after a quote change, with the NBBO mid-price marked. Row A: X posts a bid at 20.01 while Y's ask at 20.01 is stale, so the NBBO is locked at 20.01 for about 0.6 ms and the mid-price rises in two half-tick steps, from 20.005 to 20.010 to 20.015. Row B: Y's ask at 20.01 is taken by a trade, the NBBO spread widens to two ticks for tens of microseconds and the mid-price rises to 20.010, then the ask is refilled and the mid-price returns to 20.005.}]{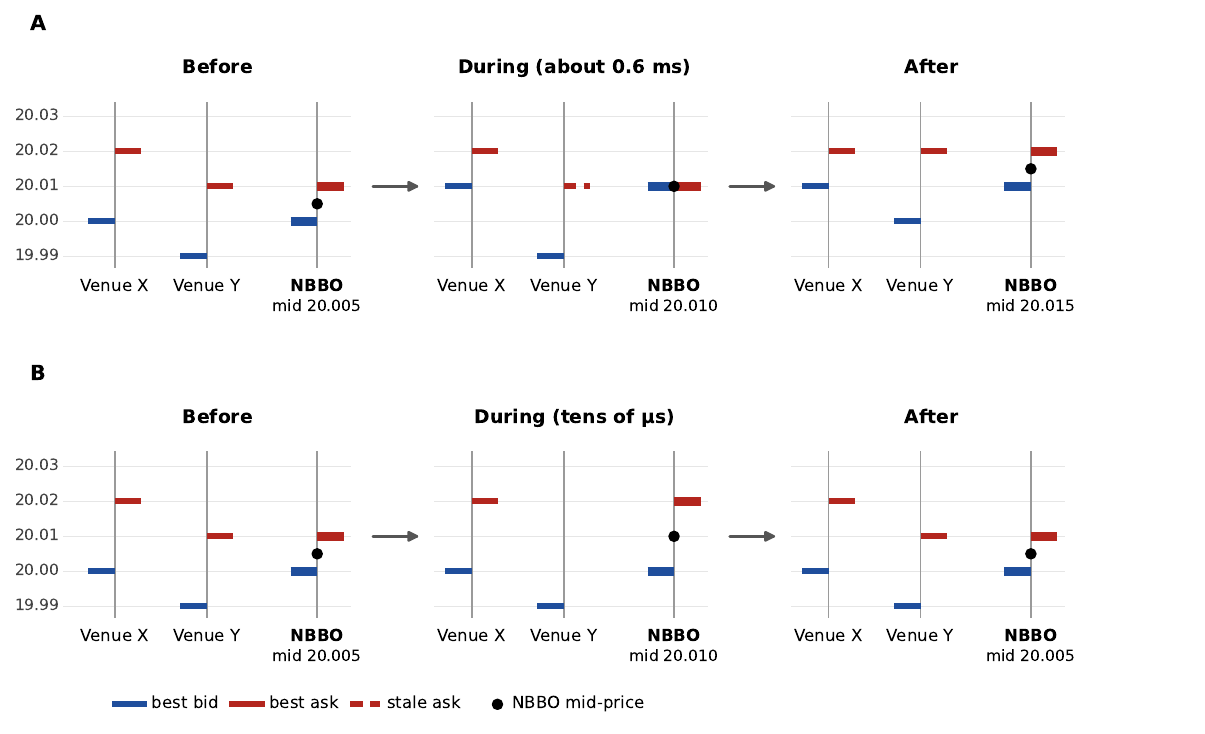}
\caption{Brief states in the consolidated NBBO (layout after \citealp{shkilko2008}, Fig.~1). Columns: each venue's best bid and ask on a one-cent ladder, and the NBBO the SIP composes from them. \textbf{A.} A one-tick move through a lock: X's new bid at 20.01 reaches the SIP before Y's updated ask, so the NBBO is locked for about $0.6\ms$. \Eo{} records two up moves of the mid, \Ei{} and \Em{} one. \textbf{B.} A brief positive-spread state: the only ask is taken by a trade and refilled tens of microseconds later. \Eo{} records an up and a down move, \Ei{} keeps both, \Em{} none. Dashed: Y's stale ask, already lifted, whose update has not reached the SIP. Prices are illustrative.}
\label{fig:brief}
\alttext{Two rows of three price-ladder diagrams. Each diagram shows the best bid and best ask on venues X and Y and the NBBO composed from them, before, during and after a quote change, with the NBBO mid-price marked. Row A: X posts a bid at 20.01 while Y's ask at 20.01 is stale, so the NBBO is locked at 20.01 for about 0.6 ms and the mid-price rises in two half-tick steps, from 20.005 to 20.010 to 20.015. Row B: Y's ask at 20.01 is taken by a trade, the NBBO spread widens to two ticks for tens of microseconds and the mid-price rises to 20.010, then the ask is refilled and the mid-price returns to 20.005.}
\end{figure}

Figure~\ref{fig:brief} shows the mechanism in a one-tick market, in the layout of \citet{shkilko2008}. Each column of the figure is a venue's best bid and best ask on a one-cent price ladder, and the last column is the NBBO that the SIP composes from them, with its mid-price. In panel A, the only ask at 20.01, on exchange Y, is lifted, and at about the same moment a trader on exchange X posts a bid at 20.01. The SIP receives X's new bid before Y's updated ask, so for a fraction of a millisecond it publishes a locked NBBO, 20.01/20.01. When Y's update arrives, the spread is one tick again, at a price one tick higher. Table~\ref{tab:example} shows the same move as a sequence of NBBO states. The mid-price moves up twice, by half a tick each time, $0.6\ms$ apart, from 20.005 to 20.010 and then to 20.015. Economically, however, it is one move of the price level, from 20.005 to 20.015.

\begin{table}[!t]
\centering
\caption{A one-tick price move in a large-tick stock as recorded in the consolidated NBBO. Times are illustrative; the lock duration is typical of 2017 SIP data.}
\label{tab:example}
\begin{tabular}{rcccl}
\toprule
Time & Best bid & Best ask & Spread & Mid-price \\
\midrule
$t$ & 20.00 & 20.01 & 1 tick & 20.005 \\
$t + 7.9\sx$ & \textbf{20.01} & 20.01 & locked & 20.010 ($\uparrow$ half tick) \\
$t + 7.9\sx + 0.6\ms$ & 20.01 & \textbf{20.02} & 1 tick & 20.015 ($\uparrow$ half tick) \\
\bottomrule
\end{tabular}
\end{table}

In liquid large-tick stocks, a move through a lock is the normal way a price moves. For Intel (INTC) over 61 Wednesdays between November 2016 and December 2017, the NBBO changes about 1,800 times a day between 10:00 and 15:30. After a change, the spread is one tick in 50--53\% of cases and locked in 36--49\%. The counts of one-tick states and of locks are nearly equal, because most moves follow the sequence one tick, locked, one tick at a new price. On May 17, 2017, the day we examine in detail, only 4\% of price moves avoid a lock. Locks last a median of $0.4$--$0.7\ms$, while one-tick states last a median of about eight seconds. The gaps between consecutive NBBO changes are therefore of two kinds: on June 5, 2017 the median gap for INTC is $1.09\ms$, but the 75th percentile is 22.6 seconds. Across the eight stocks of our sample, 41--50\% of the gaps between raw NBBO changes in May 2017 are shorter than one millisecond.

Three facts establish that these locks come from the fragmented, consolidated feed and not from the price. First, every lock involves two different exchanges. An exchange's own book cannot lock, because its matching engine executes an incoming bid at or above its own best ask instead of displaying it. In the NBBO file that carries the exchange codes of the best bid and ask, every one of 359--1,401 locks a day on six stock-days has different exchanges on the two sides.\footnote{\citet{favreau2022} drop the rare NBBO locks with one exchange on both sides as data errors. We find none.} The NBBO we rebuild from each exchange's own quotes for four stocks in May 2017 (Section~\ref{sec:robust}) confirms this on a larger scale: none of its 69,848 locks in SIP time and none of its 46,112 locks in exchange time has the same exchange on both sides. Second, lock duration does not depend on the distance between the venues involved. In the rebuilt NBBO of the same four stocks, in exchange time, the median lock between two exchanges in the same data center lasts $0.21$--$0.34\ms$, against $0.21$--$0.31\ms$ between exchanges in different data centers. In SIP time the figures are $0.24$--$0.38\ms$ against $0.24$--$0.70\ms$. Locks are therefore not the light-speed delay between buildings. Third, when we rebuild the NBBO from each exchange's own timestamps (Section~\ref{sec:robust}), locks become 20--48\% fewer and shrink to a median of about $0.25\ms$, close to one inter-venue travel time. The consolidated feed stretches each lock to its recorded $0.4$--$0.7\ms$ through its own processing and the timing of quote updates. These dislocations between the SIP's view and the exchanges' are the ones measured by \citet{ding2014} and \citet{bartlett2019}. What matters here is not their size but that each of them turns one price move into two recorded events.

Not every brief state is a lock. Some have a positive spread: the spread widens to two ticks for tens of microseconds and returns. In panel B of Figure~\ref{fig:brief}, the only ask at 20.01 is taken by a trade and refilled at the same price tens of microseconds later, so the mid moves up and back down. This can happen in two ways. Either a best quote is cancelled and replaced within microseconds with no trade (the flickering quotes of \citealp{hasbrouck2018}), or a marketable order consumes all the depth at the best price and liquidity providers refill the level. Matching these states to the TAQ trades file settles which of the two it is. For the eight stocks in May 2017, 86--99\% of sub-millisecond positive-spread states begin with a trade at the price that disappeared, within 3 ms of the state's start (82--99\% counting round lots only), against a chance level of 2--6\% and at or above the share for persistent price moves. In exchange time, for the four simulation stocks, 63--87\% of these states have a trade at that price whose participant timestamp equals the quote update's own within 10 $\mu$s (79--95\% within 100 $\mu$s), the signature of an execution that removes the displayed quote \citep{battalio2026}, against 44--77\% of persistent moves, and a coincidence within 10 $\mu$s has probability below $10^{-4}$. Excluding the trades reported off-exchange through the FINRA trade reporting facilities, 21--34\% of all trades, changes these shares by at most 0.005. They are depletion, not quote flicker. What happens within the next millisecond is one of two things. Either the depleted level is refilled and the old price returns, so the mid flickered and no persistent level was set (17--58\% of the states, most common in the stocks with the deepest books). Or the other side of the book follows, and a one-tick move is completed in two half-tick steps that the raw feed records as two events less than a millisecond apart (41--80\%, most common in the busiest stocks). In INTC such states are few. However, in the other seven stocks they leave 10--42\% of the gaps between events under one millisecond after locks are merged. For IBM, a high-priced stock with almost no locks (0.6\% of states), a third of all gaps between raw mid-price changes are under one millisecond, all of them from positive-spread states. The problem is therefore not confined to locks or to large-tick stocks.

\subsection{The Target Process and the Event Definitions}
\label{sec:target}

Each NBBO change starts a new \emph{state}, that is, an interval during which the best bid and ask are constant. For each state we record its start, its spread in ticks, its mid-price in half-ticks and its duration, until the next change or the end of the session. The process we model is defined from these states.

\begin{definition}[Target process]
Fix a persistence threshold $h > 0$ and a lock allowance $h_L \ge h$. Remove from the consolidated NBBO every locked or crossed state shorter than $h_L$ and every state shorter than $h$. The target process is the sequence of changes in the mid-price between the states that remain, each stamped at the start of the state that establishes the new level and labelled ``up'' or ``down'' by its direction.
\end{definition}

The definition is observable and reproducible, and it needs no assumption about what any trader could see. We use $h = 1\ms$ and $h_L = 5\ms$ in the main analysis, with $h = 2\ms$ and $h$ from $0.25$ to $5\ms$ as robustness checks and $h_L$ held fixed. The separate allowance for locked and crossed states is deliberate. A locked market is not a price level at all, 97\% of locks are shorter than 5 ms, and a lock that outlasts a millisecond is latency, not a persistent level (Section~\ref{sec:robust}). Without the allowance, the 1--5 ms locks of BAC would survive the merge as states and be fitted as fast excitation. Note that a state starts with an NBBO price change, so size-only updates do not start one. A change is measured between consecutive surviving states, so a short excursion and return produces no event. We examine three event series built from the same states:
\begin{itemize}
\item \Eo{} (raw): every change of the NBBO mid-price, as recorded. This is the series a Hawkes model is usually fitted to.
\item \Ei{} (locks merged): locked or crossed states shorter than $h_L = 5\ms$ are deleted, as are zero-duration states from tied timestamps. Every change in the mid-price between consecutive surviving states is an event.
\item \Em{} (all brief states merged): as \Ei, and in addition every state shorter than $h$ is deleted, whatever its spread. This is the target process and the main definition.
\end{itemize}
In Table~\ref{tab:example}, \Eo{} has two up events $0.6\ms$ apart, while \Ei{} and \Em{} have one up event, at the time the ask moves to 20.02. In panel B of Figure~\ref{fig:brief}, \Eo{} has an up and a down event. \Ei{} keeps both, because the spread stays positive, and \Em{} has none, because the state is shorter than $h$. A merged move is stamped when the brief state ends, that is, when the new level is established. A move through a brief state is therefore dated 0.4--0.7 ms after its first quote, a measurement error in event times that is small against the fastest memory of about 10 ms. The size of a move is not used, since after merging almost all moves in these stocks are of minimal size and we model timing only. Events with identical timestamps are spread evenly within one microsecond. Such ties are rare (21 of 2,509 NBBO rows for INTC on May 17, 2017) and affect raw gaps only. Merging roughly halves the event count (Section~\ref{sec:data}).

Two further claims are kept separate from the definition. The first is that \Em{} approximates the price moves of a latent price-formation process better than raw events do. This claim rests on the stability of the estimates over thresholds from 1 to $5\ms$, on the reconstruction of the NBBO in exchange time, and on the simulation study, each under stated assumptions. The second is any statement about market endogeneity, which is an interpretation of the fitted kernel and is made cautiously. Note also that the merge is retrospective: whether a state lasts $h$ is known only $h$ after it begins, so \Em{} is a reconstruction of the price path, not a real-time event stream. Finally, the definition does not say that the merged states are not trading. Locks are disagreements between venues about a price that is changing, and sub-millisecond positive-spread states are the moment after a trade takes out a quote. Both are trading activity. The definition removes them because they are not persistent price levels.

\subsection{Importance of the Definition for a Hawkes Model}
\label{sec:why}

A Hawkes model explains clustering as excitation. The only route by which past events can raise the current event rate is the kernel, since the baseline depends on time alone. Any regularity in which events follow events closely, whatever its cause, is therefore estimated as excitation: the reaction of traders to a price change, but equally the mechanical pairing of two recorded events, a common shock, or bundled timestamps. The kernel is a Granger-type notion of triggering, not a causal one \citep{eichler2017}. \citet{filimonov2015} show that the bundling of messages into packets, and the randomization of timestamps that it forces, create clustering that a Hawkes fit reads as endogeneity. A lock followed by its resolution, or a depletion followed by its refill, is mechanical pairing of exactly this kind.

When the model is fitted to raw NBBO changes, it sees that about half of all events are followed within a millisecond by another event, and it explains this with a kernel component of memory below a millisecond and weight close to one half. That component describes how the consolidated feed resolves brief states, not how traders react to price changes. However, its weight enters the branching ratio and is read as endogeneity. The link is nearly an identity. A component whose memory is shorter than the typical gap it fits integrates to about the share of events that are followed by another within that memory. The fast weight therefore tracks the share of gaps under one millisecond, which is 41--50\% in our sample and about 40\% in the fits of \citet{lee2023} to IBM and AAPL. At the same time, every real move is counted twice, which halves the weights the slower components receive per event and changes the timescales the fit attributes to them. The fitted kernel therefore mixes the mechanics of fragmented quoting with the dynamics of the price. The standard summaries of a Hawkes fit, the branching ratio above all, mix them in proportions that depend on the stock, the month and the clock.

It is well established that financial Hawkes kernels need several timescales \citep{bacry2015, lee2023}. How much of that structure depends on how the event series is built from consolidated data, and what the timescales look like once the artifacts of the feed are removed, remain open questions. The rest of the paper answers both questions. We present the model and the summaries it supports in Sections~\ref{sec:model} and~\ref{sec:analysis}, a simulation in which price moves are generated from the fitted model and the feed's artifacts are added on purpose in Section~\ref{sec:simulation}, and the panel in Section~\ref{sec:empirical}.

\section{The Model}
\label{sec:model}

In this section, we specify the bivariate Hawkes model with a refractory period that we fit to the target process, describe its estimation, and discuss the choices it involves.

\subsection{Definition}
\label{sec:definition}

Let $N_u(t)$ and $N_d(t)$ count the up and down moves of the target process on a trading session $[0, T]$, here 10:00 to 15:30, and let $\mathcal{H}_t$ be the history of both before $t$. A Hawkes process \citep{hawkes1971} specifies the conditional intensity of each type, $\lambda_i(t) = \lim_{\epsilon \downarrow 0} \epsilon^{-1} \Pr(\text{a type-}i\text{ event in } [t, t+\epsilon) \mid \mathcal{H}_t)$, as a baseline plus contributions from past events. The model we fit is
\begin{equation}
\lambda_i(t) = \mathbf{1}\{t \notin D_t\} \Big[ \mu(t) + \sum_{j \in \{u, d\}} \sum_{t^j_m < t} \phi_{ij}(t - t^j_m) \Big],
\qquad i \in \{u, d\},
\label{eq:model}
\end{equation}
with three ingredients.

\runin{Kernel.} The kernel is a sum of $K$ exponentials with decay rates shared by all four pairs of types,
\begin{equation}
\phi_{ij}(s) = \sum_{k=1}^{K} \alpha_{ij,k}\, \beta_k\, e^{-\beta_k s}, \qquad \beta_1 > \beta_2 > \dots > \beta_K > 0 .
\label{eq:kernel}
\end{equation}
Component $k$ has memory $1/\beta_k$ seconds, and $\alpha_{ij,k} \ge 0$ is the expected number of type-$i$ moves that one type-$j$ move triggers directly through that component. Up and down moves are treated as mirror images: $\alpha_{uu,k} = \alpha_{dd,k} = a_k$ (self-excitation) and $\alpha_{ud,k} = \alpha_{du,k} = c_k$ (cross-excitation, a move triggering a move in the opposite direction). The kernel then has $3K$ parameters. A sum of exponentials approximates the slowly decaying kernels found in financial data over a range of scales \citep{hardiman2013, bacry2016}, keeps the computational advantages of the exponential form, and gives each component an interpretable memory and weight \citep{lallouache2016, lee2023}. The model of \citet{leeseo2017} is the special case with $K = 1$ and a constant baseline, fitted without a refractory period and one day at a time.

\runin{Baseline.} The baseline $\mu(t)$ is piecewise constant. It takes one level in each 30-minute bin of each day, eleven bins a day, and the same level applies to up and down moves. It absorbs the intraday pattern of activity, higher near the open and the close, and the changes in activity from day to day, so that the kernel has to explain only the clustering beyond them. A constant baseline would make the intraday pattern look like self-excitation and inflate the branching ratio \citep{filimonov2015, omi2017}.

\runin{Refractory period.} The gate $D_t = \bigcup_{n:\, t_n < t} [t_n, t_n + \delta)$ is the union of the windows of length $\delta$ after the events of either type before $t$, on which the intensity is zero. We call $\delta$ the refractory period, or dead time, and these windows the dead windows. Since the gate is defined from the past only, it is predictable, and an event is never inside its own window. The target process with threshold $h$ has no two events closer than $h$, by construction. Without the gate, the model would give an event a positive chance right after another, and the fastest component would be fitted to a region of the time axis that contains no events. We therefore set $\delta = h$. Note that the gate is part of the model, so the likelihood accounts for it exactly (Section~\ref{sec:estimation}).

One kernel is shared by all trading days of a calendar month, and the baseline levels are specific to each day and bin. Each day starts with no excitation carried over from the previous one. A month of a core stock has 10,000--40,000 clean events, enough to pin down a component with memory of about 30 seconds, while monthly fits still show how the estimates move across the 14 months of the sample. For a month of 22 days with $K = 4$ the model has $22 \times 11 = 242$ baseline levels, 8 excitation parameters and 4 decay rates.

\runin{Branching representation.} A Hawkes process can be built as a population \citep{hawkesoakes1974, daley2003}. Immigrant events arrive as a Poisson process with intensity $\mu(t)$. Each event of type $j$ produces a Poisson number of offspring of type $i$ with mean $\alpha_{ij} = \int_0^\infty \phi_{ij}(s)\,ds$, at delays with density $\phi_{ij}/\alpha_{ij}$. With the symmetric kernel the branching matrix is $\mathbf{A} = \sum_k \big(\begin{smallmatrix} a_k & c_k \\ c_k & a_k \end{smallmatrix}\big)$, with eigenvalues $\sum_k (a_k + c_k)$ and $\sum_k (a_k - c_k)$, so the branching ratio, the spectral radius of $\mathbf{A}$, is
\begin{equation}
n = \sum_{k=1}^{K} (a_k + c_k),
\label{eq:br}
\end{equation}
and $a_k + c_k$ is the excitation carried by timescale $k$. We call $n$ the total branching ratio. Without the gate and with a constant baseline, the process is stationary when $n < 1$ \citep{hawkes1971, bremaud1996}. In that case $n$ is the expected fraction of events that are offspring, not immigrants, which is why \citet{filimonov2012} proposed it as a measure of endogeneity. The model we fit is neither ungated nor stationary. Its baseline changes by bin and by day, and the gate bounds the event rate by $1/\delta$ whatever the value of $n$, so no stability condition is needed and none is imposed (no clean fit exceeds 0.79 and no raw fit 0.83). Under the gate, $n$ is the spectral radius of the ungated kernel, and the offspring interpretation is approximate, by at most 1\% on these data (Appendix~\ref{app:mass}). In addition, the gated process has fewer events than the ungated one built from the same driving randomness, since the gate only removes them (Appendix~\ref{app:likelihood}). In Section~\ref{sec:analysis}, we explain why we do not use $n$ as the summary of a fit, and what we use instead.

\subsection{Likelihood and Estimation}
\label{sec:estimation}

The log-likelihood of a multivariate point process observed on $[0, T]$ is
\begin{equation}
\ell = \sum_{i \in \{u,d\}} \Big( \sum_{n:\, \text{type } i} \log \lambda_i(t_n) - \int_0^T \lambda_i(s)\, ds \Big),
\label{eq:loglik}
\end{equation}
summed over the days of the month. For exponential kernels, the sum over the past in \eqref{eq:model} is computed recursively. Let
\begin{equation}
R_{j,k}(t) = \sum_{t^j_m < t} e^{-\beta_k (t - t^j_m)}
\label{eq:decayed}
\end{equation}
be the decayed count of past type-$j$ events seen by component $k$. Outside the dead windows, the intensity is $\lambda_i(t) = \mu(t) + \sum_j \sum_k \alpha_{ij,k} \beta_k R_{j,k}(t)$, and the decayed counts update from one event to the next,
\begin{equation}
R_{j,k}(t_n) = e^{-\beta_k (t_n - t_{n-1})} \big( R_{j,k}(t_{n-1}) + \mathbf{1}\{t_{n-1} \text{ is of type } j\} \big),
\label{eq:recursion}
\end{equation}
so the intensity at all $N$ events costs $O(KN)$ operations \citep{ozaki1979}. The compensator has a closed form, given in Appendix~\ref{app:likelihood}, with three parts: the integral of the baseline over each bin, less its overlap with the dead windows; for each past event, the mass of its kernel up to $T$; and, subtracted, the mass that falls in the dead windows of later events. The last part is a sum over events of $R_{j,k}(t_n^+)\,(1 - e^{-\beta_k \delta_n})$, where $\delta_n$ is the length of the dead window after event $n$, cut at the next event, and $R_{j,k}(t_n^+)$ includes the event at $t_n$. We verified the expression against a brute-force computation and against numerical integration of the gated intensity.

For fixed decay rates $\boldsymbol\beta$, the intensity at each event is linear in the remaining parameters $\boldsymbol\theta = (\mu, a, c)$, and the compensator is linear in them too. Hence \eqref{eq:loglik} is a sum of logarithms of linear functions minus a linear function, which is concave in $\boldsymbol\theta$ and is maximized reliably by a bounded quasi-Newton method with $\boldsymbol\theta \ge 0$. This gives the profile likelihood $\ell^*(\boldsymbol\beta) = \max_{\boldsymbol\theta} \ell(\boldsymbol\theta \mid \boldsymbol\beta)$, which we maximize over $\log \boldsymbol\beta$ by the Nelder--Mead simplex method. Note that the profile likelihood is not concave in $\boldsymbol\beta$ and can have several local maxima. In one stock-month, for instance, two triples of timescales are local optima 63 log-likelihood units apart. We therefore start the simplex from a full grid of initial decay-rate combinations (for $K = 4$, the 15 combinations of four rates from a grid of six, plus three ladders of typical fitted values) and keep the best optimum. This is the multi-start search that \citet{filimonov2015} show to be necessary.

We choose the number of components by the Bayesian information criterion, $\mathrm{BIC} = -2\ell + p \log N$ with $p$ the number of parameters and $N$ the number of events, compared across $K$ for the same event series. Likelihoods are not comparable across event definitions. Compared between $K = 3$ and $K = 4$, with the baseline parameters common to both, BIC prefers $K = 4$ in 103 of the 112 clean stock-months. The exceptions are BAC in eight months (February to October 2017), whose third scale is bimodal across months and carries little excitation, and INTC in January 2017. We therefore use $K = 4$ for every series and $K = 3$ for BAC, the one stock whose months mostly prefer it. A fifth component changes nothing we report. In May 2017 BIC prefers $K = 5$ over $K = 4$ in five of the eight stocks, but the fifth component lands between 2 ms and 10 s, at scales the four already cover, or carries no weight (BAC), and $n(1\sx)$ and $n(10\sx)$ move by at most 0.003 and 0.006.

Table~\ref{tab:spec} collects the specification of every fit reported in the paper. Unless stated otherwise, the baseline has one level per day and 30-minute bin. The clean fits start from the full grid. The raw fits start from the default grid plus two points built from the three-component fit of the same stock-month, which reach the full-grid optimum in all eight May stock-months to the last decimal of the log-likelihood.

We assess goodness of fit with the time-rescaling theorem. If the model is correct, the increments of the compensator between consecutive events of a type are independent standard exponential variables, so their Kolmogorov--Smirnov distance from that distribution is the test statistic. Because the parameters are estimated, the usual critical values are too lenient. We therefore calibrate the test by a parametric bootstrap: we simulate from the fitted model by Ogata's thinning method with the dead windows, refit, and compare the observed statistic with the refits' \citep{cavaliere2023, kling2026}. The same bootstrap gives model-based standard errors and a bias estimate for every parameter. In Section~\ref{sec:robust}, we add a block bootstrap over trading days, which treats days as independent blocks, as the likelihood does, and does not assume that one kernel generated every day. The block bootstrap is the primary measure of uncertainty for the horizon masses, while the parametric one measures estimation noise under the fitted model only.

\begin{table}[!t]
\centering
\caption{Specification of the fits reported. $K$: kernel components; $\delta$: refractory period; $h$: persistence threshold; $h_L$: lock allowance. A dash in the $h$, $h_L$ column means that the series applies no threshold of that kind: raw events merge nothing, and \Ei{} merges locks only.}
\label{tab:spec}
\scriptsize
\setlength{\tabcolsep}{3pt}
\begin{tabular}{llcccl}
\toprule
Series & Stocks & $K$ & $\delta$ & $h$, $h_L$ & Reported in \\
\midrule
\Em{} (clean, main) & all & 4$^\ast$ & 1 ms$^\dagger$ & 1 ms$^\dagger$, 5 ms & Tables~\ref{tab:main}--\ref{tab:regime}; bootstraps; simulation \\
\Em{} (robustness) & all & 4$^\ast$ & $h$ & 0.25--5 ms, 5 ms & Table~\ref{tab:threshold}; Section~\ref{sec:robust} \\
\Eo{} (raw) & all & 4 & 0 & ---, --- & Tables~\ref{tab:horizon}, \ref{tab:simraw}, \ref{tab:blockboot}, \ref{tab:priceunits}, Figure~\ref{fig:priceunits} \\
\Ei{} (locks merged) & all & 3 & 0 & ---, 5 ms & Table~\ref{tab:horizon}; simulation \\
Lee (2023) replication & IBM, AAPL & 2 & 0 & 1 ms, 5 ms (clean) & Table~\ref{tab:ibm}; constant baseline; daily fits \\
\bottomrule
\multicolumn{6}{l}{$^\ast$ BAC: $K = 3$. \quad $^\dagger$ BAC: 2 ms.}
\end{tabular}
\end{table}

\subsection{Discussion of the Model}
\label{sec:discussion}

In this subsection, we discuss four choices in the model: the refractory period, the symmetric kernel, the baseline and the pooling of days, and the number of components.

\subsubsection{The refractory period}
The gate removes the part of the time axis on which the target process cannot have events. Its effect on the estimates is confined to the fastest component. With the gate, the fastest memory is shorter by a median 24\% across the panel, and by 43--56\% in TGT and ATVI, because without it the fit stretches that component to cover the empty first millisecond after each event. The effective branching ratio differs by a median 0.0003 with and without the gate, and the kernel masses within one and ten seconds by amounts of the same order.

Note that the gated model is an approximation to the law of the target process, not its exact description. The persistence filter deletes both members of a genuine pair of moves closer than $h$ and merges same-direction pairs into one event, which a dead time does not. For a component with memory 10 ms, about $1 - e^{-0.1} \approx 10\%$ of its offspring fall within 1 ms of their parent, so the approximation matters most for the fastest component. The simulation of Section~\ref{sec:simulation}, which passes an ungated true process through the real filter, measures the consequence as a bias of 3--9\% in the fastest memory and 3--7\% in $n(1\sx)$. Kernel mass inside an event's own dead window is never realized, so the quantities we report are defined from the dead time onward (Section~\ref{sec:mass}). In addition, the gate follows every event, so an event's kernel is also switched off in the dead windows of later events, and the expected number of realized offspring is slightly smaller than the kernel summaries say. On the actual event histories of May 2017 the difference is below 0.001 for the core stocks and 0.005 for TGT, about 1\%, and we neglect it.

\subsubsection{One kernel for both directions}
The symmetric kernel treats an up move after an up move exactly as a down move after a down move, and a reversal in either direction alike. A linear kernel with $a_k, c_k \ge 0$ cannot lower the intensity of same-direction moves below the baseline plus their self-excitation, so neither the symmetric nor the full form can represent the excess of reversals reported in Section~\ref{sec:misfit}. The monthly counts of up and down moves are balanced to within a few percent. As a robustness check, we also fit a kernel with separate excitation for each of the four pairs of directions. It moves the lack of fit between directions without removing it, and BIC does not prefer it. The symmetric form has half the kernel parameters and one summary per timescale, which is what the comparison across event definitions needs.

\subsubsection{The baseline and the slowest component}
A baseline per day and 30-minute bin is a choice, and the choice interacts with the slowest kernel component: clustering on the scale of minutes is baseline when the baseline is fine and excitation when it is coarse. In Section~\ref{sec:tradeoff}, we show that the total branching ratio moves by a factor of up to 1.7 across baseline resolutions on which AIC and BIC disagree, while kernel mass within one second does not move and within ten seconds moves only at the finest baselines. Pooling the days of a month is the other choice. Fitted one day at a time, as in \citet{lee2023}, the likelihood of a four-component kernel is multimodal at 500--4,000 events a day and the slow component is not identified. A month pins it down. The cost is that the kernel is assumed common to the days of the month. We test this assumption with the day-block bootstrap and relax it, in part, in Section~\ref{sec:activity}.

\subsubsection{Interpretation of the four components}
The fitted memories of a four-component kernel, about 10 ms, 100 ms, 2 s and 30 s in the core stocks, are a representation of the distribution of response times, not four mechanisms. A smooth kernel decaying as a power law over the same range is represented by a few exponentials with roughly equal weights. The simulation of Section~\ref{sec:simulation} shows that a four-exponential fit to such a kernel recovers its cumulative mass within 5\% from 100 ms to 10 s. Accordingly, what we interpret are cumulative quantities (the kernel mass within a horizon), not the individual components, whose decay rates are estimated with relative standard errors of 2--20\%.

\section{What the Estimates Measure}
\label{sec:analysis}

In this section, we fix the quantities that the empirical study reports and explain, with short derivations, why the standard summary of a Hawkes fit is the wrong one for comparing event definitions.

\subsection{Kernel Mass within a Horizon under a Refractory Period}
\label{sec:mass}

For the model of Section~\ref{sec:model}, the kernel mass between the dead time and a horizon $H$ is
\begin{equation}
n(H) = \sum_{k=1}^{K} (a_k + c_k)\big( e^{-\beta_k \delta} - e^{-\beta_k H} \big), \qquad H > \delta,
\label{eq:nH}
\end{equation}
a truncated kernel integral, with $\delta = 0$ for raw events. Without the refractory period, $n(H)$ is the expected number of moves, up or down, that one move triggers directly within time $H$. As $H \to \infty$, it tends to the branching ratio \eqref{eq:br}. With the refractory period, the mass inside an event's own dead window is never realized, so the realized excitation of an isolated event is $n(\infty) = \sum_k (a_k + c_k) e^{-\beta_k \delta}$. We call this quantity the effective branching ratio. Because the gate follows every event, an event's kernel is also switched off inside the dead windows of later events, so the expected number of realized offspring is slightly below \eqref{eq:nH}. In Appendix~\ref{app:mass}, we show that on the actual event histories the difference is at most 1\%, and we treat \eqref{eq:nH} as the summary.

We report $n(1\sx)$ and $n(10\sx)$ for every fit, for three reasons that the rest of the paper substantiates. First, they are cumulative, so they do not depend on how the response-time distribution is split into components: $K = 4$ and $K = 5$ give the same values within 0.001 in the simulation and within 0.006 on the data (Sections~\ref{sec:simulation} and~\ref{sec:estimation}), and a four-exponential fit to a power-law kernel recovers them within a few percent (Section~\ref{sec:simulation}). Second, they are the part of the kernel that the data identify. The fast components show no parametric-bootstrap bias and have relative standard errors of 2--18\% on their memories, whereas the slowest component trades off against the baseline (Section~\ref{sec:tradeoff}). Under the refractory period, the quantity a clean fit estimates is the kernel mass on $(h, H]$, since mass below $h$ is removed by the definition and cannot be recovered (Section~\ref{sec:simlimits}). We state the clean estimand as $n(H)$ on that window and, where raw and clean fits are compared, report the raw mass on the same window as well as from zero (Section~\ref{sec:rawclean}). Third, they are the quantities whose distortion by the event definition is largest and cleanest, since the artifact of Section~\ref{sec:nbbo} lives below one millisecond. Note that $n(H)$ is not baseline-free by construction, because the fitted kernel can change with the baseline. Empirically, it hardly does at these horizons.

Two caveats apply to $n(H)$. First, it is a per-event quantity, so a comparison across event definitions compares different event units (a raw event is often half of one price move). In Section~\ref{sec:rawclean}, we restate the comparison in price units. Second, it is a kernel summary under the fitted model. What it says about the market is an interpretation, which we make in Section~\ref{sec:activity} with the caveats of Section~\ref{sec:misfit}.

\subsection{What the Raw Branching Ratio Measures}
\label{sec:identity}

The horizon masses are the summary we report. We now ask what the standard summary, the branching ratio, measures when the fit is to raw events. A counting argument gives the answer. Suppose that the target process has $N_m$ events in a month and branching ratio $n_m$, and that the raw series adds to some of these moves a second event within a millisecond (the resolution of a brief state), for $N_0$ raw events in all. Let $s$ be the share of raw events that are followed by another within a millisecond, which we measure as the share of gaps under $1\ms$. The fit explains the paired events with its fastest component: each second event is read as an offspring of the first, so per raw event that component carries about $s$. The slower components see the real excitation, but diluted. A clean move triggers on average $n_m$ later clean moves, and in the raw series each of those begins with one raw event that the slow components can claim, while the parents are now spread over $N_0$ raw events instead of $N_m$. Per raw event the slow weight is therefore about $n_m N_m / N_0$. Adding the two, we obtain
\begin{equation}
n_0 \;\approx\; s + n_m \frac{N_m}{N_0}.
\label{eq:identity}
\end{equation}
Across the 112 stock-months of the panel, \eqref{eq:identity} predicts the raw branching ratio with median error $-0.02$ and median absolute error 0.035 (Section~\ref{sec:rawclean}), and the median of $N_m/N_0$ is 0.46. The raw branching ratio is therefore about one half artifact (the term $s \approx 0.45$--$0.50$) plus the real excitation diluted by about one half. Two consequences follow. First, raw ratios bunch together, between 0.62 and 0.83 across stock-months, while clean ratios range from 0.27 to 0.79, so raw data compress the differences between stocks and months. Second, the ``30\% overstatement'' of the total that a naive comparison reports is the net of an additive artifact and a multiplicative dilution. On fixed horizons, where the excitation is identified, the artifact dominates and the ratio is three (Section~\ref{sec:rawclean}).

The fastest component's weight is nearly an identity with the data. A component whose memory is shorter than the gaps it fits integrates to about the share of events followed by another within that memory. Hence the finding that raw-event fits attribute 40--50\% of events to a sub-millisecond component, ours and \citeauthor{lee2023}'s (\citeyear{lee2023}), shows that the component captures those gaps. It does not by itself show what the gaps are. The evidence that the removed states are artifacts therefore has to come from their properties (Section~\ref{sec:nbbo}), from the stability of the estimates across thresholds and clocks (Section~\ref{sec:robust}), and from the simulation (Section~\ref{sec:simulation}), not from the fit.

\begin{remark}
If every removed event were within $1\ms$ of its partner, $s$ would equal $1 - N_m/N_0$ and \eqref{eq:identity} would say that the number of immigrant events, $N(1-n)$, is the same under both definitions. This ``immigrant invariance'' does not hold in the data: raw fits attribute about 30\% more events per day to the baseline than clean fits (median 1.31 across the 112 stock-months, 1.07--1.59 by stock, with equal numbers of kernel components). The excess has an exact counterpart in the data. It is the removed raw events that are not sub-millisecond followers, that is, the entries into brief states that resolve more than a millisecond later (locks of 1--5 ms, merged under the lock allowance) and the entries into round trips in which a level flickers and returns. Counted on the May 2017 states, these are 3--16\% of raw events, almost all of them long locks (13\% for BAC, 6--7\% for INTC and T, 1--5\% in the busy stocks), with round trips 0.1--1.3\%, against an excess of 2--9\% of raw events in the fits. The two agree where the kernel is well identified (INTC: 0.060 against 0.064) and differ in two known directions. In BAC and T, the raw fit's sub-millisecond tail absorbs part of the long-lock resolutions as offspring, which is the relation's error for BAC seen from the other side. In the busy stocks, both fits put slow-component mass into the baseline, the raw fit more. The simulation of Section~\ref{sec:simulation} reproduces both the relation's errors and the excess of raw baseline events from the brief-state mechanism alone.
\end{remark}

\subsection{The Baseline and the Slowest Component}
\label{sec:tradeoff}

The relation explains why the raw branching ratio is not comparable with the clean one. The clean total has a difficulty of its own, which comes from the baseline and not from the feed. The total branching ratio \eqref{eq:br} is dominated by the slowest component, and the slowest component is the one the baseline competes with. Table~\ref{tab:baseline} shows the main specification for INTC in May 2017 refitted with finer and coarser baselines. Each coarsening lengthens the slowest memory and raises the total, from 0.44 with 30-minute bins to 0.74 with one level per day. The parametric bootstrap bias of the total shrinks at the same time, from $-0.06$ to $-0.01$, so the bias with fine baselines is itself a symptom: part of a 30--50 second clustering is absorbed by the bins. In the other direction, with 10- and 5-minute bins, the slowest memory shortens (46 s, 18 s, 5 s) and the total falls (0.44, 0.27, 0.18). The bias is $-0.06$ at 10-minute bins and $-0.03$ at 5-minute bins, where the slowest memory of 5 s sits well inside the bin and is identified again. Clustering on the scale of minutes is baseline when the baseline is fine and excitation when it is coarse, and the information criteria do not settle the matter. We repeated the six fits for all eight stocks in May and November 2017. In every one of the 16 stock-months the total rises with the bin length, from 0.33--0.70 with 30-minute bins to 0.64--0.88 with one level per day. Among 30-minute, 60-minute and daily baselines, AIC prefers 30-minute bins in all 16 and BIC one level per day in 14. Over all six resolutions, AIC prefers 10- or 5-minute bins in 9 of the 16, and BIC one level per day in 11 and 110-minute bins in 5.

\begin{table}[!t]
\centering
\caption{INTC, May 2017, \Em{} with $K = 4$ and the refractory period, under six baseline resolutions. Total: effective branching ratio (Section~\ref{sec:mass}), which is 0.003 below \eqref{eq:br} here. Bias: parametric bootstrap, 30 replicates (99 for the 30-minute row; Monte Carlo standard error 0.001--0.002). $n(1\sx)$ and $n(10\sx)$ from the dead time.}
\label{tab:baseline}
\small
\begin{tabular}{lccccccc}
\toprule
Baseline bins & AIC & BIC & Slowest memory & Total & Bias & $n(1\sx)$ & $n(10\sx)$ \\
\midrule
5 min & 146,723 & 157,957 & 5 s & 0.18 & $-0.033$ & 0.100 & 0.17 \\
10 min & 146,551 & 152,214 & 18 s & 0.27 & $-0.059$ & 0.102 & 0.19 \\
30 min (main) & \textbf{146,539} & 148,488 & 46 s & 0.44 & $-0.059$ & 0.103 & 0.21 \\
60 min & 146,550 & 147,655 & 64 s & 0.51 & $-0.046$ & 0.103 & 0.22 \\
110 min & 146,649 & 147,248 & 83 s & 0.59 & $-0.029$ & 0.103 & 0.22 \\
one per day & 146,865 & \textbf{147,126} & 134 s & 0.74 & $-0.013$ & 0.103 & 0.23 \\
\bottomrule
\end{tabular}
\end{table}

The total is therefore specification-dependent, a sharper form of the point made by \citet{filimonov2015} about baselines and apparent criticality. The dependence is sensitivity under the specifications examined, not a proof of formal non-identifiability. However, it means that a total branching ratio reported without its baseline specification is not a number that can be compared across papers, and that the question whether the market is ``near criticality'' cannot be settled by it. The fixed-horizon masses behave differently. $n(1\sx)$ is 0.100--0.103 across all six resolutions of Table~\ref{tab:baseline}, and across the 16 stock-months it moves by at most 0.011 from 5-minute bins to one level per day, including four November stock-months in which the coarsest baselines rearrange the timescales. $n(10\sx)$ moves by at most 0.027 from 30 minutes upward, but falls by 1--11\% at 10-minute bins and by 3--22\% at 5-minute bins, where the ten-second horizon starts to share the minute-scale clustering with the baseline. Our reporting rule follows. $n(1\sx)$ is stable under every baseline examined and $n(10\sx)$ from 30-minute bins upward, so we keep the 30-minute baseline as the main specification, report both masses with standard errors, and give the total only with its specification stated. Note that the rule rests on an identifying assumption that should be stated: scale separation, under which exogenous variation in the intensity is slower than the baseline bin and the endogenous response faster than tens of seconds, so that the two can be told apart by their timescale. Stability across the specifications examined is evidence for it, not a proof. The fall of $n(10\sx)$ at 5-minute bins shows where the assumption weakens, and the bursts of Section~\ref{sec:misfit} show where it fails.

This section fixes what the empirical study reports and why. First, the kernel mass within one and ten seconds, on the window from the threshold, is the part of the kernel that the data identify and that the baseline leaves alone. Second, the raw branching ratio is an artifact term plus the real excitation diluted by the doubled event count, so it is not comparable with the clean one. Third, the total branching ratio depends on the baseline specification and is reported only together with it. The simulation of Section~\ref{sec:simulation} tests the first two claims on price moves generated from the fitted model, and the empirical study of Section~\ref{sec:empirical} applies all three to the data.

\section{Simulation Study}
\label{sec:simulation}

Merging removes sub-millisecond gaps by construction, so the absence of a sub-millisecond component under \Em{} proves nothing by itself, and a fit to real data cannot say how far its estimates are from the true values. In this section, we generate price moves from the fitted model, put them through the observation mechanism of Section~\ref{sec:nbbo}, calibrated on the data, and ask what each event definition and the estimator recover. The logic is that of \citet{leeseo2023}'s simulation study, with one addition: between the true process and the estimate sits the real event pipeline, applied to a synthetic consolidated feed.

\subsection{Design}
\label{sec:simdesign}

\runin{Data-generating process.} The true process is the fitted \Em{} model of May 2017 for four stocks that span the sample, INTC, MDLZ, BAC and TGT: $K = 4$, the May baselines for every day and bin, and the May kernel. The true process has no refractory period, so it can generate events closer than one millisecond. The dead time enters only when \Em{} is fitted. For each replicate, we simulate a month of 22 days by Ogata's thinning method.

\runin{Observation layer.} We then turn each clean move into a path of consolidated-NBBO quotes with the brief states of Section~\ref{sec:nbbo}. For a move from a one-tick spread, we draw the path as one of four types with probabilities calibrated on the stock's May data: direct (both quotes moving at once); through a lock (the bid rising to the old ask first and the ask following after a locked interval, two NBBO changes each moving the mid half a tick, as in Figure~\ref{fig:brief}A); through a widening (the ask rising first and the bid following); or a half-tick move of one quote. We add flickers, brief states that return to the previous quotes (Figure~\ref{fig:brief}B), at the calibrated rate per clean move, split between locks and widenings in the calibrated proportion. We resample the durations of brief states from the empirical distributions of the stock-month. A brief state never starts before the previous NBBO change, which is how close real moves interact in the consolidation. We keep the layer simple on purpose, with one brief state per move and no long locks. How far it reproduces the real raw-event statistics is itself one of the results. The real pipeline then builds \Eo, \Ei{} and \Em{} from the synthetic quotes, and we run the fits of Section~\ref{sec:empirical} on every replicate: \Em{} with $K = 4$ and the dead time, \Em{} without the dead time, \Eo{} with $K = 4$ and \Ei{} with $K = 4$, each started from the full grid of Section~\ref{sec:estimation} with the true decay rates added as one more start.

\runin{Experiments.} We run eight experiments, 1,000 replicate-months and 5,200 fits in all, none of which failed:
\begin{enumerate}
\item \emph{base}: the calibrated brief states (50 replicates per stock);
\item \emph{share0} and \emph{share50}: no brief states, and half the calibrated probabilities and flicker rate (25 each);
\item \emph{long}: the calibrated states with durations tripled, a feed with a longer lock tail (25);
\item \emph{subms}: the true kernel plus a genuine sub-millisecond component, memory $0.3\ms$ and mass 0.10, with the calibrated brief states (25); and \emph{subms-clean}, the same process with no brief states at all (25);
\item \emph{power law}: a true kernel of the form $\phi(t) \propto (c + t)^{-1-\theta}$ with the May total mass, $(c, \theta)$ fitted to the May cumulative kernel mass on the log scale ($c$ from 2.8 to 13.5 ms, $\theta$ from 0.06 to 0.23 across stocks), represented by 15 exponentials from 0.2 ms to 200 s and fitted, as always, with four (25);
\item \emph{scaled baselines}: the May kernel with the May baselines multiplied by 0.5, 0.71, 1, 1.41 and 2, ten replicates at each scale, so that activity varies fourfold with the kernel fixed (50).
\end{enumerate}
The base experiment adds fits with $K = 3$ and $K = 5$ and fits pooled over 10, 5 and 1 days. The two sub-millisecond experiments add \Eo{} with $K = 5$. Throughout, the fitted $n(H)$ is the kernel mass between the dead time and $H$, as in \eqref{eq:nH}, and the true $n(H)$ is the true mass from 0 to $H$, so that any mass the dead time removes counts against the estimate.

\subsection{Recovery and the Raw-Event Distortion}
\label{sec:simrecovery}

The base experiment asks two questions: how close the clean fit comes to the kernel that generated the moves, and whether the measured brief states alone produce the raw-event distortion of Section~\ref{sec:identity}. Table~\ref{tab:simrecovery} answers the first. \Em{} recovers the kernel mass within one second and within ten seconds to within 3--7\% of the true values, with standard deviations across replicates of 0.002--0.006. Without any brief states the bias of $n(1\sx)$ is $-2$ to $-7\%$, and a fit to the true event times themselves recovers it within 1\% in every stock. This part of the bias therefore comes from the merge, which removes the genuine moves closer than one millisecond (0.4--6\% of the events), and not from the estimator. The total branching ratio is biased down by 5--20\% in INTC, MDLZ and BAC, through the slowest component, whose fitted memory is 0.59--0.85 of the true one. The fit to the true event times shows the same bias of the total (1--21\%), so this one is the estimator's. The parametric bootstrap reports the same bias on the real data (Section~\ref{sec:robust}). In TGT, whose slowest memory is 10 s and is recovered (ratio 0.93), the total is within 2\%. Note that the bias of the total depends on the stock in a way that the fixed-horizon masses do not, which is the empirical content of Section~\ref{sec:tradeoff}. The refractory period matters for the fastest memory only. Without it that memory is overstated by 26--57\%, with it by 3--9\%, while $n(1\sx)$ and $n(10\sx)$ are the same within 0.005 either way. $K = 4$ and $K = 5$ give the same masses within 0.001, and $K = 3$ differs by up to 0.015.

\begin{table}[!t]
\centering
\caption{Recovery of the kernel masses and the total by \Em{} ($K = 4$, $1\ms$ refractory period) in the base experiment. True: the value for the kernel that generated the moves, with $n(H)$ taken from 0 to $H$. \Em: median over 50 replicate-months, standard deviation across replicates in parentheses, and the median's deviation from the true value in percent. The fitted $n(H)$ is taken from the dead time to $H$. Total: the branching ratio \eqref{eq:br}.}
\label{tab:simrecovery}
\small
\begin{tabular}{lcccccc}
\toprule
Stock & True $n(1\sx)$ & \Em{} $n(1\sx)$ & True $n(10\sx)$ & \Em{} $n(10\sx)$ & True total & \Em{} total \\
\midrule
INTC & 0.106 & 0.101 (0.003), $-5\%$ & 0.215 & 0.206 (0.005), $-4\%$ & 0.440 & 0.376, $-15\%$ \\
MDLZ & 0.238 & 0.228 (0.003), $-4\%$ & 0.364 & 0.353 (0.006), $-3\%$ & 0.498 & 0.476, $-5\%$ \\
BAC  & 0.085 & 0.080 (0.002), $-5\%$ & 0.160 & 0.152 (0.005), $-5\%$ & 0.405 & 0.325, $-20\%$ \\
TGT  & 0.443 & 0.413 (0.003), $-7\%$ & 0.571 & 0.549 (0.004), $-4\%$ & 0.633 & 0.646, $+2\%$ \\
\bottomrule
\end{tabular}
\end{table}

Table~\ref{tab:simraw} answers the second question and is the main result of the simulation. With price moves generated from the fitted model and nothing added but the measured brief states, the raw-event fits reproduce what the real raw-event fits show. The raw/clean ratio of kernel mass within one second is 5.3, 2.6, 6.4 and 1.6 against 5.4, 2.8, 8.0 and 1.8 in the data, in the same order across stocks, and the raw sub-millisecond mass is 0.31--0.43 against 0.39--0.49. Raw $n(1\sx)$ reaches 86--97\% of its real-data value. The shortfall is the layer's. It records 1.8--2.0 raw NBBO changes per clean move, against 2.1--2.3 in the data, because a real move sometimes passes through more than one brief state, a lock followed by a widening or two locks as several venues update, and the layer allows one. The gap is therefore one-sided by construction and largest for BAC, whose lock chains are the longest. The raw slowest memories are 15--68\% of the true value, as in the real raw fits with the same $K$, and merging locks only (\Ei) still inflates $n(1\sx)$ by 36--140\%.

The counting relation \eqref{eq:identity} holds in the simulation with errors of the size seen in the data (Table~\ref{tab:simraw}, fourth column). The errors are within $\pm 0.02$ for MDLZ and TGT in both, and BAC's deviation in the data, $+0.14$, comes out as $+0.17$, so the simulation reproduces where the relation fails and by about how much. INTC is the one case in which the simulation errs ($+0.07$) and the data do not. In addition, the raw fits attribute more events to the baseline than the clean fits, by 2--18\% in the simulation against 7--30\% in the data (last column). The mechanism reproduces the direction of this excess and, except in INTC, most of its size. The failure of immigrant invariance noted in Section~\ref{sec:identity} is therefore what the brief-state mechanism produces, not evidence against it. Both definitions put too many events in the baseline, clean fits by 2--14\% and raw fits by 13--24\%, which is the counterpart of the slow-component bias.

\begin{table}[!t]
\centering
\caption{Raw-event distortion in the simulation (base experiment, medians over 50 replicate-months) and in the data (May 2017; raw: \Eo{} with $K = 4$, clean: the main specification). Columns: raw over clean kernel mass within one second and within ten seconds; raw kernel mass within the first millisecond; error of the relation \eqref{eq:identity}, the raw branching ratio minus its prediction; and events attributed to the baseline per day, raw fit over clean fit. Each cell gives the simulated value, then the data value.}
\label{tab:simraw}
\footnotesize
\setlength{\tabcolsep}{3pt}
\begin{tabular}{lccccc}
\toprule
& Raw $\div$ clean $n(1\sx)$ & Raw $\div$ clean $n(10\sx)$ & Raw mass below $1\ms$ & Relation error & Baseline, raw $\div$ clean \\
Stock & simulation / data & simulation / data & simulation / data & simulation / data & simulation / data \\
\midrule
INTC & 5.3 / 5.4 & 2.84 / 2.92 & 0.36 / 0.40 & $+0.07$ / $0.00$ & 1.01 / 1.22 \\
MDLZ & 2.6 / 2.8 & 1.90 / 1.99 & 0.43 / 0.48 & $0.00$ / $-0.03$ & 1.18 / 1.29 \\
BAC  & 6.4 / 8.0 & 3.71 / 4.36 & 0.30 / 0.39 & $+0.17$ / $+0.14$ & 1.07 / 1.07 \\
TGT  & 1.6 / 1.8 & 1.38 / 1.46 & 0.42 / 0.49 & $-0.01$ / $-0.02$ & 1.17 / 1.30 \\
\bottomrule
\end{tabular}
\end{table}

Three further results from the base experiment bear on the empirical study. First, model order. BIC chooses $K = 4$ in every replicate for INTC, MDLZ and TGT. For BAC it chooses $K = 3$ in 48 of 50 replicates although the true kernel has four components, because BAC's two slow components, 13 s and 74 s, are not separable in a month. The real data show the same preference for $K = 3$. Second, pooling. Fitting 10 days, 5 days or a single day adds no bias to $n(1\sx)$ and $n(10\sx)$ beyond the month's. The standard deviation of $n(1\sx)$ grows from 0.002--0.003 for a month to 0.010--0.017 for a day, while the effective branching ratio from a single day has standard deviation 0.02--0.07. A day suffices for the short-horizon masses, and the slow end needs a month. Third, goodness of fit under a correct model. The KS statistic of \Em{} is 0.004--0.009 at the median and 0.006--0.013 at the 95th percentile. These values are the yardstick against which we judge the real misfit in Section~\ref{sec:misfit}.

\subsection{Sensitivity and Limits}
\label{sec:simlimits}

The base experiment holds the brief states and the kernel at their May 2017 values. Each conclusion above could depend on that choice. In this subsection, we vary the share and the durations of the brief states, add a genuine sub-millisecond component to the true kernel, change the kernel's shape, and scale the baselines.

\runin{The share of brief states.} With none, half or all of the calibrated brief states, \Em's $n(1\sx)$ is the same within 0.003, while the raw inflation grows with the share (INTC: 1.03, 3.8 and 5.3 times). The clean estimates do not depend on how much of the artifact the feed contains. The raw ones are the artifact.

\runin{Durations longer than the threshold.} With the brief-state durations tripled, states that outlast the one-millisecond threshold survive the merge and contaminate \Em: $n(1\sx)$ is 17\% too high for INTC, 4\% for MDLZ, 5\% for TGT, which acquires a spurious 0.7 ms component, and 180\% for BAC (0.238 against 0.085). The threshold must therefore exceed the brief-state durations. For the actual durations it does, as the threshold curve of Section~\ref{sec:robust} shows, and the exchange-time reconstruction identifies the one stock, BAC, whose SIP lag approaches the threshold.

\runin{A genuine sub-millisecond component.} With a $0.3\ms$ component of mass 0.10 added to the true kernel, two things happen. When the feed has no brief states, raw fits with $K = 4$ identify the component exactly, with memory 0.30 ms and mass 0.100 in every stock, and recover $n(1\sx)$ within 0--3\%. \Em{} removes it by construction: its $n(1\sx)$ is 15--51\% below the true value, almost exactly the 0.10 removed, and it recovers the kernel beyond one millisecond within $+8\%$ ($+13\%$ for TGT), slightly inflated because 10--18\% of the genuine events are merged away. When the feed has the calibrated brief states, neither definition recovers it. \Em{} erases the component, and the raw fits' sub-millisecond mass rises by only 0.05--0.06 of the 0.10 added, with a fifth component failing to isolate it, because the brief states put a component of mass 0.3--0.4 at the same timescale. A genuine sub-millisecond reaction is therefore recovered by the fit from raw events when there is nothing to clean, and it is lost in both definitions once brief states are present, which in consolidated data they always are. The claim we make about the sub-millisecond scale is correspondingly limited: not that no such reaction exists, but that consolidated data cannot identify one, and that what raw-event fits find there is the feed.

\runin{A power-law kernel.} When the true kernel decays as a power law, a four-exponential \Em{} fit recovers its cumulative mass from one millisecond within 5\% from 100 ms to 10 s (within 13\% at 10 ms and 10\% at 100 s). The fitted memories are about 6--25 ms, 40--180 ms, 0.3--1.3 s and 4--12 s, carrying roughly equal masses, which is the shape of a discretised power law. The total is 5--17\% low, because the tail beyond the slowest fitted memory is absorbed by the 30-minute baseline, the mechanism of Section~\ref{sec:tradeoff} again. The misspecification is invisible to the KS statistic (median 0.004--0.010, against 0.004--0.009 under the correct model), and the raw-event distortion is the same under this data-generating process (raw/clean $n(1\sx)$ 1.5--2.7). The horizon masses, not the components, are what the estimator delivers under either kernel shape.

\runin{Activity and excitation.} With one kernel and baselines scaled over a fourfold range of activity, $n(1\sx)$, $n(10\sx)$, the effective branching ratio and the slowest memory are flat. The slope of $n(1\sx)$ per doubling of activity is at most 0.002 in absolute value (INTC: 0.104, 0.103, 0.104, 0.100, 0.100 at the five scales), and the correlations with log events per day lie between $-0.23$ and $+0.22$. In the data, $n(1\sx)$ rises with activity across months, within the day and across days (Section~\ref{sec:activity}), by amounts far outside the replicate spread, so the estimator does not produce the relation.

\runin{Scope of the result.} The simulation does not show that the real brief states arose from the simulated mechanism. It shows that the mechanism, calibrated on their measured properties, produces what the raw-event fits find, and that the clean definition recovers the true kernel under it. The evidence that the real states are of this kind is in Section~\ref{sec:nbbo}.

Three results carry into the empirical study. First, the clean fit recovers the kernel mass within one and ten seconds, so clean $n(1\sx)$ and $n(10\sx)$ can be read as estimates, while the total cannot, because of the slow-component bias. Second, the measured brief states alone produce the raw-event distortion, so a raw/clean gap of the size seen in the simulation is the mark of the feed, not of the price. Third, a correctly specified model sets the yardstick for the misfit. In Section~\ref{sec:empirical}, we report the raw and clean fits on the data, check the clean estimates against the threshold, the clock and the baseline, and measure the misfit against this yardstick.

\section{Empirical Study}
\label{sec:empirical}

In this section, we describe the data, compare raw and clean events on the summaries of Section~\ref{sec:analysis}, check the robustness of the clean estimates, examine how excitation moves with activity, and report where the model still fails.

\subsection{Data}
\label{sec:data}

\runin{Sample.} The data are the NYSE Trade and Quote database (Millisecond TAQ) through Wharton Research Data Services (WRDS), with microsecond timestamps from 2016 \citep{taq2017}. We build the event series from the completed NBBO table (\texttt{complete\_nbbo}), which adds the quotes the standard NBBO file omits when one exchange holds both the best bid and the best ask \citep{holden2014, favreau2022}. We use the standard file, which carries the exchange codes and the exchanges' own timestamps, for the venue and latency diagnostics. The period is October 31, 2016 to December 29, 2017, 294 trading days, beginning after a format change in the TAQ files in late October 2016. We drop the single October day, since the fits are by calendar month, and three early-close days, leaving 290 days in the fourteen months fitted. None of the eight stocks was in the Tick Size Pilot of 2016--2018, which covered small-capitalization stocks. Each day is observed from 10:00 to 15:30, avoiding the open and the close, as in \citet{leeseo2017}.

\runin{Stocks.} We chose the stocks in two stages: the 40 most traded US common stocks priced between \$10 and \$60 in 2017, from the CRSP daily stock file \citep{crsp2017}, and then a tick-size profile of each on three days. The eight chosen span the range from fully large-tick to intermediate. INTC, MU and MDLZ (Nasdaq-listed) and BAC, WFC and T (NYSE-listed) sit at a one-tick spread at least 99.8\% of the time on the screening days, with 35--46\% of their NBBO changes locking or crossing, while TGT and ATVI spent 91\% and 67\% of the screening days at one tick. We add IBM and AAPL in January 2017 for the comparison with \citet{lee2023}. We saved the NBBO price changes on WRDS as one file per stock-day and transferred them out. All event construction and fitting happens outside WRDS, so an event definition can change without a new query.

\runin{Tick regime.} We classify every stock-month by time at a one-tick spread, weighted by duration and excluding locked and crossed states. By this rule, 80 of the 112 stock-months are large-tick (at least 99\%): every month of INTC, BAC, WFC and T, 11 of 14 for MU (96.8--98.7\% in three months) and 10 of 14 for MDLZ (98.0--98.8\% in four). TGT and ATVI cross regimes within the sample as their prices move. TGT is small-tick from November 2016 to February 2017 (\$65--76), intermediate (90--99\%) or large-tick from March to August 2017 (\$54--56, large-tick in April and May), and small-tick again from September 2017 (\$58--63), down to 38\% of time at one tick in December. ATVI is large-tick in November 2016 (\$40), intermediate up to \$59, and small-tick from July 2017 (\$60--64; 86\% of time at one tick in July and 33--55\% after). The sample thus holds 80 large-tick, 18 intermediate and 14 small-tick stock-months. The last two groups come mostly from TGT and ATVI, with the seven MU and MDLZ months just below the large-tick line among the intermediate ones. We state the headline results for the large-tick months, with TGT and ATVI as the contrast, and return to what changes with the regime in Section~\ref{sec:activity}.

\runin{Event statistics.} Table~\ref{tab:events} gives the three event series for May 2017. Raw NBBO changes number 27,000--126,000 a month, of which 41--50\% are followed by another within a millisecond. Merging locks alone (\Ei) removes 17--49\% of the events, a third to a half in the core stocks, but leaves 4--42\% of the gaps under a millisecond, which are the positive-spread brief states of Section~\ref{sec:briefstates}. Merging every state shorter than a millisecond (\Em) halves the raw count and leaves, by construction, none. The median gap between clean events is 4--16 seconds in the core stocks and about a second in TGT and ATVI. After merging, consecutive clean moves go in the same direction 42--48\% of the time, so clean price moves show mild reversal.

\begin{table}[!t]
\centering
\caption{Event statistics, May 2017 (22 trading days, 10:00--15:30). Time at one tick includes locked and crossed states in the time base here; the regime classification above excludes them.}
\label{tab:events}
\small
\begin{tabular}{lrrrrrrrr}
\toprule
& INTC & MU & MDLZ & BAC & WFC & T & TGT & ATVI \\
\midrule
\% time at 1 tick & 99.9 & 99.9 & 99.7 & 99.9 & 99.8 & 99.9 & 99.0 & 95.1 \\
\% states locked or crossed & 47.9 & 45.3 & 40.3 & 42.3 & 33.8 & 37.4 & 14.5 & 28.9 \\
\midrule
\Eo{} events & 32,710 & 67,863 & 49,517 & 34,894 & 87,135 & 26,508 & 126,480 & 121,806 \\
\Eo: \% gaps $< 1\ms$ & 45.1 & 50.3 & 50.1 & 41.4 & 47.5 & 49.1 & 50.4 & 49.9 \\
\Ei{} events & 16,646 & 34,560 & 28,422 & 17,867 & 54,612 & 14,652 & 105,490 & 84,466 \\
\Ei: \% gaps $< 1\ms$ & 4.3 & 10.4 & 18.1 & 16.2 & 26.3 & 25.0 & 42.2 & 29.4 \\
\Em{} events & 15,888 & 30,607 & 22,715 & 14,898 & 39,761 & 10,849 & 58,818 & 57,068 \\
\Em: median gap (s) & 10.9 & 5.2 & 5.4 & 12.1 & 4.2 & 15.7 & 1.0 & 1.3 \\
\bottomrule
\end{tabular}
\end{table}

\subsection{Estimation Results: Raw against Clean Events}
\label{sec:rawclean}

Whether the sub-millisecond component is the price's or the feed's, and what the raw branching ratio measures, are settled in this subsection. We ask whether the component survives the clean definition, first in the specification of \citet{lee2023} and then across the panel, how much kernel mass the two definitions place within fixed horizons and on a common window, what the difference amounts to in price units, and whether the counting relation of Section~\ref{sec:identity} describes the panel.

\subsubsection{The sub-millisecond component and where it goes}
We begin where the literature stands. \citet{lee2023} fits raw NBBO mid-price changes of IBM and AAPL one day at a time, with a constant baseline, and finds a fast kernel component with decay rate of 1,000--2,000 per second to which about 40\% of events are attributed. We pulled the same two stocks for January 2017, inside his sample period, and fitted his specification to the raw series and to the clean one (Table~\ref{tab:ibm}). The replication is close. For IBM on January 3, 2017, the raw fit gives a fastest memory of $1.18\ms$ against his $1.14\ms$, with 41\% of events attributed to it against the 44\% he reports for IBM over 2017. On the clean series the sub-millisecond component is gone in both stocks, and the fastest memory becomes about 9 ms for IBM and 20 ms for AAPL (Table~\ref{tab:ibm}), carrying a smaller but substantial share of events. IBM is the instructive case. It has almost no locks (0.6\% of states), and its sub-millisecond states are positive-spread brief states, 32\% of all gaps between raw events, which merging locks alone leaves untouched and \Em{} removes. The problem is therefore not confined to locks or to large-tick stocks.

In the benchmark specification, cleaning leaves the branching ratio nearly unchanged (IBM 0.61 against 0.60, AAPL 0.70 against 0.63), and the counting relation of Section~\ref{sec:identity} overpredicts the raw ratio by 0.07--0.10. The reason is the model's capacity. With two components, the raw fit spends one on the brief states and has one for everything slower, which carries 0.17--0.18 of mass beyond two milliseconds where the dilution of the clean excitation predicts 0.31--0.40. With three components, the raw ratios rise to 0.72 and 0.82, the relation's error falls to $-0.03$ and $-0.05$, and the raw mass beyond two milliseconds approaches the diluted value. The benchmark's branching ratio is invariant to cleaning because its raw fit trades the real slow excitation for the brief-state component one for one. The contamination of the branching ratio described in Section~\ref{sec:identity} is a statement about fits with room for both.

\begin{table}[!t]
\centering
\caption{Daily fits in the specification of \citet{lee2023} (one day at a time, constant baseline, $K = 2$), January 2017, 20 days: medians [10th--90th percentile]. Share of events: event attribution to the fastest component, as in that paper.}
\label{tab:ibm}
\small
\begin{tabular}{lccc}
\toprule
& Fastest memory & Share of events from it & Branching ratio \\
\midrule
IBM, \Eo & 1.10 ms [0.65--1.45] & 41\% [38--45] & 0.61 [0.56--0.65] \\
IBM, \Em & 8.8 ms [7.5--11.3] & 29\% [25--33] & 0.60 [0.53--0.65] \\
AAPL, \Eo & 0.38 ms [0.31--0.43] & 52\% [48--53] & 0.70 [0.65--0.76] \\
AAPL, \Em & 19.8 ms [15.5--25.1] & 22\% [20--28] & 0.63 [0.57--0.72] \\
\bottomrule
\end{tabular}
\end{table}

\begin{figure}[!t]
\centering
\includegraphics[width=\linewidth, alt={Two panels, INTC and TGT. Each shows the fitted kernel on raw events and on clean events as t times phi of t against time since the event on a log axis from 10 microseconds to 100 seconds. The raw kernel is a single peak below one millisecond; the clean kernel is zero below one millisecond and spreads its mass between 10 milliseconds and 100 seconds.}]{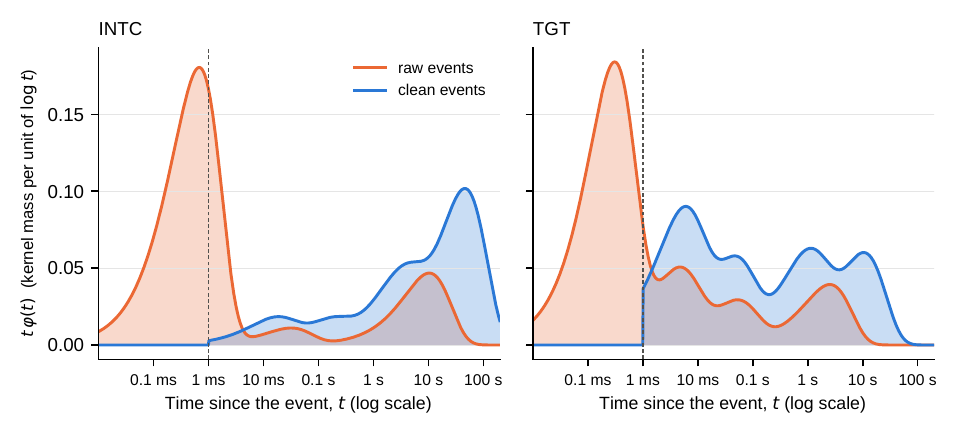}
\caption{Fitted kernel on raw and clean events, INTC and TGT, May 2017. Raw: \Eo{} with $K = 4$; clean: \Em{}, main specification. $\phi = \phi_{uu} + \phi_{ud}$ is the rate of moves in either direction triggered by one up move, drawn as $t\,\phi(t)$ on a log time axis, so that the area under a curve between two times is the kernel mass there. The dashed line marks one millisecond, the merge threshold and refractory period of the clean fit.}
\label{fig:kernel}
\alttext{Two panels, INTC and TGT. Each shows the fitted kernel on raw events and on clean events as t times phi of t against time since the event on a log axis from 10 microseconds to 100 seconds. The raw kernel is a single peak below one millisecond; the clean kernel is zero below one millisecond and spreads its mass between 10 milliseconds and 100 seconds.}
\end{figure}

The same holds in every stock and month of our panel. Fitted to raw events, the model places 0.39--0.49 of its kernel mass below one millisecond (May 2017, eight stocks), in one or two components with memories from 0.03 to 0.7 ms, and its branching ratio is 0.67--0.78. Across the 112 stock-months, that sub-millisecond mass tracks the share of raw gaps under a millisecond, as Section~\ref{sec:identity} says it must. Two features of the raw data separate this pairing from a fast component of a Hawkes process. First, the number of further raw events within a millisecond of an event is one far more often than a Poisson count with the same probability of at least one would give. The share with two or more is 6--13\% against 12--16\% in six stocks, half or less for INTC, and matches Poisson only in BAC and T, whose locks chain with widenings. Second, the followers are half-tick quote updates (full-tick moves are under 2\% of them), 59--87\% in the direction of the event they follow: the second half of a move, not a new one. On clean events no sub-millisecond component remains. Figure~\ref{fig:kernel} draws the two fitted kernels for INTC and TGT in May 2017, as $t\,\phi(t)$ on a log time axis, so that the area under a curve between two times is the kernel mass there. The raw kernel is a spike below one millisecond. The clean kernel has nothing there, and its mass lies between 10 ms and 100 s. The fastest memory is 8--13 ms in five of the six core stocks, 5 ms in MDLZ and 3--4 ms in TGT and ATVI, and the branching ratio is 0.32--0.56 in the core stocks and about 0.65 in TGT and ATVI (Section~\ref{sec:activity} gives the full panel). Part of this result is by construction, since \Em{} has no gaps under a millisecond. The informative part is what replaces the component and how the rest of the kernel changes, which we measure in the next two subsections.

\subsubsection{Kernel mass within one and ten seconds}
Table~\ref{tab:horizon} compares the two definitions on the summaries of Section~\ref{sec:mass}, over all 112 stock-months. Raw-event fits put 3.1 times the clean kernel mass within one second and 2.0 times within ten seconds, and every stock-month is above one (ranges 1.4--8.7 and 1.2--5.5). The total branching ratio, raw 33\% above clean, understates the difference by a wide margin, because the slow component that dominates the total is sensitive to the baseline under both definitions. Three quarters of the raw one-second mass, 0.46 of 0.63, lies in the first millisecond, which is the resolution of brief states. Raw data also compress the differences between stocks: clean $n(1\sx)$ ranges from 0.10 (BAC) to 0.45 (TGT) across stocks, a factor of 4.5, whereas raw ranges from 0.57 to 0.70, a factor of 1.2. Merging locks alone (\Ei, fitted with $K = 3$) removes only part of the distortion, 0.40 within one second against 0.20.

The last two rows of each block put the two definitions on a common window. From the threshold $h$ onward, the raw mass is 0.71 of the clean mass within one second and 0.64 within ten seconds. From 5 ms onward it is 0.48 and 0.50, which is the dilution $N_m/N_0 = 0.46$ of \eqref{eq:identity} (the ratio of raw to diluted clean mass is 1.03 [0.96--1.10] within one second and 1.08 [1.05--1.11] within ten). Beyond 5 ms, then, the raw fit sees the real excitation diluted by the doubled event count and nothing else. Between $h$ and 5 ms the raw fit sees more than the diluted excitation: on $(h, 1\sx]$ the raw mass is 1.5 times the diluted clean mass (1.9--2.9 for INTC, WFC, T and BAC, 1.1--1.2 for TGT and ATVI), which is the tail of the brief states, the locks of 1--5 ms that the lock allowance removes from the clean series and the tail of the sub-millisecond component.

\begin{table}[!t]
\centering
\caption{Kernel mass within 1 s and 10 s and the total branching ratio, 112 stock-months: median [interquartile range]. Clean: \Em{}, main specification. Raw: \Eo{} with $K = 4$; \Ei{} with $K = 3$.}
\label{tab:horizon}
\small
\begin{tabular}{lccc}
\toprule
& $n(1\sx)$ & $n(10\sx)$ & Total \\
\midrule
Clean (\Em), from $h$ & 0.20 [0.14--0.34] & 0.34 [0.24--0.48] & 0.54 [0.42--0.65] \\
Raw (\Eo), from 0 & 0.63 [0.60--0.68] & 0.70 [0.66--0.76] & 0.72 [0.69--0.77] \\
\quad of which in the first millisecond & 0.46 [0.43--0.48] & & \\
Locks merged only (\Ei), from 0 & 0.40 [0.27--0.56] & 0.51 [0.38--0.62] & \\
\midrule
Raw $\div$ clean, raw from 0 & \textbf{3.1} [2.0--4.2] & \textbf{2.0} [1.5--2.7] & 1.33 [1.19--1.56] \\
Raw $\div$ clean, both from $h$ & 0.71 [0.58--1.13] & 0.64 [0.57--0.87] & \\
Raw $\div$ clean, both from 5 ms & 0.48 [0.44--0.51] & 0.50 [0.48--0.53] & \\
\bottomrule
\end{tabular}
\end{table}

The comparison does not depend on the raw model's complexity. We fit the raw panel with $K = 4$, started from the default grid and from two points built on the $K = 3$ fit of the same stock-month. Refitting it with $K = 3$ gives a median raw/clean ratio of 3.08 within one second against 3.06 with $K = 4$, and 2.00 against 2.03 within ten seconds, with no stock-month moving by more than 0.15 within one second. With four components, the raw fit spends two of them below 5 ms in 104 of the 112 stock-months, on the brief states and their tail. Its slowest memory reaches the clean value only where the clean kernel is simplest, in 12 of BAC's 14 months, and stays at 2--8 s elsewhere (median 1.6--7.8 s by stock, against 27--40 s on clean events). Nor does the comparison depend on the clean model's complexity: clean $n(1\sx)$ and $n(10\sx)$ are the same with $K = 4$ and the refractory period as with $K = 3$ and none (median differences 0.001 and 0.004).

\subsubsection{The comparison in price units}
A raw event is often half of one price move, so Table~\ref{tab:horizon} compares per-event summaries of two different event units. We use two checks to put both definitions in the same unit, the mid-price, for May 2017 and the four simulation stocks, with the raw fits at $K = 4$. First, raw events move the mid by 0.50--0.53 ticks on average and clean events by 0.93--0.96, and the sizes are nearly constant within each definition. Hence $n(H)$ is also the triggered price movement in ticks per tick of price movement. Within one second it is 0.55--0.71 ticks on raw events against 0.08--0.40 on clean ones, the same ratio as per event.

Second, and more telling, we compare each fitted model with the price path itself, on the same functional for both. The data side is the mean signed change of the actual NBBO mid over $(t, t + H]$ after each event of that definition, with a standard error from the 22 daily means, since events within a day are dependent. The model side is the same statistic computed on five months simulated from the fitted model on the same days and baselines, each simulated event moving the mid by the definition's mean size. Note that the model side differs from the cascade of one isolated event, because an event inside a cluster also inherits the continuation of its ancestors' other offspring: by 23--43\% for the raw fits and by under 0.005 ticks for the clean ones. Figure~\ref{fig:priceunits} gives both from 10 ms to 10 s, and Table~\ref{tab:priceunits} in Appendix~\ref{app:priceunits} the values at 100 ms, 1 s and 10 s with their standard errors.

After a raw event the mid does continue, by $+0.09$ to $+0.21$ ticks, which is the second half of a move through a brief state. However, the raw model implies a continuation of $+0.29$ to $+0.59$ ticks, 2.7 to 3.1 times what occurs within 100 ms and within one second and 2.5 to 4.1 times within ten seconds, in every stock, a gap of ten to over a hundred standard errors. The clean model reproduces the observed path within 0.021 ticks up to one second in all four stocks, including the mild reversal of clean moves (TGT within 0.01 at every horizon), and within two standard errors at one second in three of them. The differences that remain are where the residual diagnostics of Section~\ref{sec:misfit} put them. At one second, BAC's clean model implies $+0.023$ against $+0.006$ (standard error 0.005), and at ten seconds the model understates the reversal in INTC and BAC ($-0.06$ against $-0.09$ and $-0.02$ against $-0.06$, standard errors 0.014 and 0.015), consistent with the downward bias of its slowest component. In a common unit, then, raw-event fits attribute three times as much triggered price movement within one second and imply close to three times the post-event continuation that is observed, while the clean model is within a few hundredths of a tick of the observed path.

\begin{figure}[!t]
\centering
\includegraphics[width=0.85\linewidth, alt={Four panels, INTC, BAC, MDLZ and TGT. Each shows the signed mid change after an up move, in ticks, against the horizon on a log axis from 10 milliseconds to 10 seconds. Solid lines are the observed values with a two-standard-error band and dashed lines the model. The raw model lies far above the raw observations in every stock; the clean model lies on the clean observations.}]{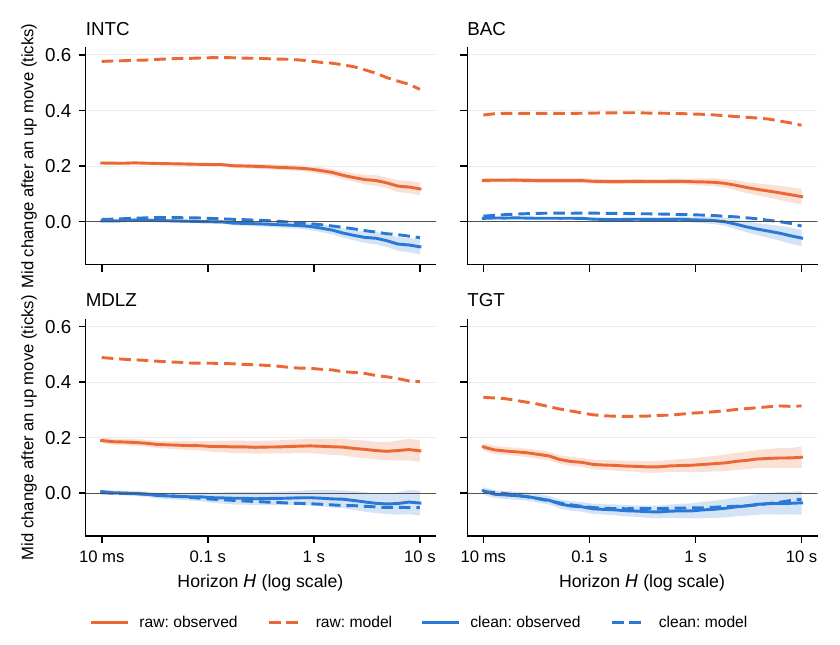}
\caption{Signed change of the mid-price after an up move, in ticks, May 2017, by horizon. Solid: observed in the data, with a band of two standard errors from the daily means. Dashed: implied by the fitted model, computed as the same statistic on five months simulated from it. Raw: \Eo{} with $K = 4$; clean: \Em{}, main specification.}
\label{fig:priceunits}
\alttext{Four panels, INTC, BAC, MDLZ and TGT. Each shows the signed mid change after an up move, in ticks, against the horizon on a log axis from 10 milliseconds to 10 seconds. Solid lines are the observed values with a two-standard-error band and dashed lines the model. The raw model lies far above the raw observations in every stock; the clean model lies on the clean observations.}
\end{figure}

\subsubsection{The counting relation in the data}
Finally, the relation \eqref{eq:identity} describes the panel. With $s$ the share of raw gaps under a millisecond and the counts of the two series, it predicts the raw branching ratio with a median error of $-0.02$ and a median absolute error of 0.035 across the 112 stock-months, with four components on both series; 76 of the 112 errors are within $\pm 0.05$. The one systematic exception is BAC, whose raw ratio is a median 0.15 above the prediction, and which the simulation reproduces (Table~\ref{tab:simraw}). BAC has the longest SIP lag of the eight stocks, and part of its locks outlast the merge threshold. The raw branching ratio is, as the relation says, the share of brief states, about one half, plus the real excitation diluted by about one half. Across stocks and months, the share of brief states is the part that varies.

\subsection{Robustness}
\label{sec:robust}

The results of Section~\ref{sec:rawclean} rest on the clean definition, so we ask here how far they depend on its free choice, the threshold $h$, with the lock allowance $h_L$ held at 5 ms, and on the clock the NBBO is built with. The last part fixes the uncertainty of a month's estimate. In this subsection, we check the clean estimates against the merge threshold, the clock, and a coarse alternative filter, and we compare two bootstraps.

\subsubsection{The merge threshold}
The threshold $h$ is the one free choice in the definition of the target process. Table~\ref{tab:threshold} refits the eight stocks in May and November 2017 with $h$ from $0.25$ to $5\ms$, with the refractory period equal to $h$ throughout. Below one millisecond the brief states come back. At $h = 0.25\ms$, 14 of the 16 stock-months again have a fastest memory of $1.3\ms$ or less, and the slow end shrinks toward the raw-event ladder (TGT, May: slowest memory 10 s to 2.3 s). At $0.5\ms$ the sub-millisecond component returns in 6 of 16. From one to five milliseconds the estimates are stable: the medians of the three slower memories move by at most 5\% and of the effective branching ratio by at most 0.013, the latter because genuine moves closer than $h$ are merged too. On the window from 5 ms, which every threshold leaves intact, the kernel mass within one and ten seconds moves by at most 3\% and 4\% at the median across the five thresholds (last two columns of Table~\ref{tab:threshold}). One millisecond is therefore the smallest threshold at which no sub-millisecond component remains, and above it the results do not depend on the threshold, except for the fastest memory. Refitting the whole panel at $h = 2\ms$ confirms the stability month by month. The effective branching ratio changes by a median $-0.005$ (largest $-0.036$) and the three slower memories by a median 9\%, 3\% and 2\%, while the fastest memory moves by more than 20\% in 39 of the 41 stock-months in which it is under 6 ms at $h = 1\ms$ (median ratio 1.70) and in 10 of the 71 in which it is not. Hence the reporting rule of Table~\ref{tab:main}: a fastest memory under about 6 ms is not resolved. Its excitation is not sensitive to $h$ between 1 and 5 ms, since the branching ratio and the horizon masses do not depend on the threshold there, but its timescale is set by the cleaning, not by the market. The rule applies to MDLZ, TGT and ATVI in most months and to the busiest months of the other stocks.

\begin{table}[!t]
\centering
\caption{Threshold curve: \Em{} with merge threshold and refractory period $h$, $K = 4$, eight stocks in May and November 2017; the lock allowance $h_L = 5\ms$ is held fixed, so $h$ varies the merge of positive-spread states and genuine moves only. Medians over the 16 stock-months of the ratio to the value at $h = 1\ms$ (events, memories, and the kernel masses on $(5\ms, H]$) and of the difference (effective branching ratio).}
\label{tab:threshold}
\footnotesize
\setlength{\tabcolsep}{3.5pt}
\begin{tabular}{lcccccccc}
\toprule
$h$ & Events & Fast & Second & Third & Slow & Eff.\ BR minus value at $1\ms$ & $n(1\sx)$ from 5 ms & $n(10\sx)$ from 5 ms \\
\midrule
0.25 ms & 1.08 & 0.09 & 0.22 & 0.39 & 0.56 & $+0.004$ & 0.96 & 0.95 \\
0.5 ms & 1.02 & 0.54 & 0.80 & 0.92 & 0.92 & $0.000$ & 0.99 & 0.99 \\
1 ms & 1 & 1 & 1 & 1 & 1 & 0 & 1 & 1 \\
2 ms & 0.98 & 1.23 & 1.05 & 1.03 & 1.02 & $-0.004$ & 1.01 & 1.02 \\
5 ms & 0.95 & 1.15 & 1.04 & 1.04 & 1.04 & $-0.013$ & 1.03 & 1.04 \\
\bottomrule
\end{tabular}
\end{table}

\subsubsection{Whose clock: an exchange-timestamp reconstruction}
The brief states of Section~\ref{sec:nbbo} are partly the SIP's own doing, so a natural question is whether the results depend on whose clock the NBBO is built with. TAQ's per-exchange quote table records each exchange's own timestamp. Following the exchange-time construction of \citet{holden2023}, without their location-specific latency adjustment, we rebuilt the NBBO from every exchange's regular quotes in two orders, by SIP time and by exchange time, for the four simulation stocks in May 2017 and again in November 2017, the busiest month of the sample.\footnote{We avoid calling the exchange-time version ``latency-free'': the standard the exchanges' clocks were held to, synchronization to within 100 $\mu$s of the NIST clock under Section 6.8 of the Consolidated Audit Trail plan, a standard the Commission tightened from the plan as filed when it approved it in November 2016 \citep{sec2016cat}, is of the same order as the exchange-time lock durations, and the reconstruction still contains the travel time between venues; ``one inter-venue travel time'' is a reading of those durations, not a measurement.} The SIP-time rebuild reproduces the completed NBBO table within a few percent in May and within 0.03\% in \Em{} counts in November, which validates the construction. In exchange time, locks are fewer and shorter but not gone: 20--48\% fewer, with median duration $0.18$--$0.31\ms$, about one inter-venue travel time. The brief states change form and do not disappear. A quote that locked the market in SIP time often arrives after the opposite side has moved, which leaves a brief two-tick spread instead. Brief widenings therefore rise by a factor of two to seven, and the share of raw gaps under one millisecond does not fall (0.47--0.55 in exchange time). The exchange-time reconstruction is thus not a threshold-free substitute for \Em.

\Em{} itself is nearly invariant to the clock. Its event counts differ by at most 2.3\% across the three versions in May and by 0.02--0.3\% in November (BAC 2.1\%). Matching events one-to-one across the two clocks, same direction within 5 ms, 97--99.7\% of \Em{} events have a counterpart, with the same size in 97--99.7\% of pairs, and the exchange-time event leads by a median 0.2--1.0 ms, the SIP's own delay. Raw events match less well (84--98\%), because brief locks in one clock become brief widenings in the other. The fitted models are the same across clocks for INTC, MDLZ and TGT in November, with $n(1\sx)$, $n(10\sx)$ and the total equal to three decimals, and nearly so in May. The three slower components, the total and $n(10\sx)$ agree in every case. The one exception is the fastest component where the SIP lag is longest. For BAC, which has the longest lag of the eight stocks (median 0.75 ms) and the largest share of SIP-time locks longer than a millisecond (0.28, against 0.11 in exchange time), $n(1\sx)$ is 18--28\% lower in exchange time, because latency-made locks that outlast the threshold survive the merge in SIP time and are counted as fast excitation. INTC shows a smaller version in May (0.103 against 0.093) and none in November. Raw-event estimates, by contrast, change with the clock throughout. The sub-millisecond memory shrinks (INTC 0.66 to 0.23 ms in May; the 40 $\mu$s fastest raw memory of INTC and MDLZ in November becomes 0.15 ms), and the slow memories move. The clean-event model beyond its fastest component therefore does not depend on whose clock is used, while raw-event estimates do. Where the SIP lag approaches the threshold, as for BAC, $n(1\sx)$ carries a clock caveat and the 2 ms threshold is the safer choice.

\subsubsection{A coarse alternative: the 0.1-second grid}
\citet{lee2024} removes the ultra-high-frequency activity of raw NBBO data by sampling the mid-price every 0.1 s and recording one event, at the time of the last change, whenever the sampled mid differs from the previous sample. His target is daily volatility, for which the filter is harmless by design. Applied to the May 2017 data of the four simulation stocks, the grid also removes the brief states: 88--96\% of its events are full-tick moves and no sub-millisecond component remains. However, the grid removes much more. Moves within one grid interval are netted into one, so the grid has 5\% fewer events than \Em{} in INTC and BAC, 16\% fewer in MDLZ and 32\% fewer in TGT, and only 2--7\% of its gaps are under 0.1 s. Fitted with our model, kernel mass within one second is 37--60\% lower than on \Em{} (INTC 0.065 against 0.103, BAC 0.034 against 0.078, MDLZ 0.113 against 0.226, TGT 0.162 against 0.403) and within ten seconds 17--39\% lower. The 10 ms and 40--130 ms components, which Table~\ref{tab:threshold} shows to be stable for thresholds from 1 to 5 ms, are largely gone. The grid also creates an artifact of its own, moves on either side of a grid boundary microseconds apart, on which the fit spends a component with memory 1--3 ms. The ordering \Eo{} $>$ \Em{} $>$ grid holds in all four stocks. A duration rule removes the brief states while keeping the clustering at 10--100 ms, whereas a sampling grid removes both.

\subsubsection{Uncertainty: parametric and day-block bootstrap}
Two bootstraps give the uncertainty of the estimates, and they answer different questions. The parametric bootstrap of Section~\ref{sec:estimation}, with 99 replicates per stock for May 2017, describes the estimator when the fitted model generates the data. Under it the fast components are unbiased and precise: the two fastest memories return within 2\% with relative standard errors of 2--18\% and their excitation within 0.0005, and $n(1\sx)$ and $n(10\sx)$ have standard errors of 0.002--0.003 and 0.004--0.005 with no measurable bias. The slowest component is biased down, by about 20\% in the core stocks (slowest memory 25--50 s) and 4--7\% in TGT and ATVI (10--18 s), with the lost excitation going to the baseline. The effective branching ratio has a bias of $-0.01$ to $-0.06$, two to five times its standard error of 0.005--0.012 in the core stocks. In November 2017 the same bootstrap gives standard errors of 0.002--0.004 for $n(1\sx)$ and 0.003--0.005 for $n(10\sx)$, again without measurable bias, and a bias of $-0.004$ to $-0.033$ in the effective branching ratio, smaller than in May because the slowest memories are shorter in November (7--31 s). The simulation reproduces this bias (Table~\ref{tab:simrecovery}), and the baseline trade-off of Section~\ref{sec:tradeoff} explains it.

The parametric bootstrap assumes that one kernel generated every day of the month. The day-block bootstrap drops that assumption. We resample trading days in blocks of five consecutive days with replacement, use the same resampled days for \Eo{} and \Em, and refit both models on each of 100 replicates, for every stock-month of the panel on the clean series and for May 2017 on both.\footnote{Each refit starts from the month's decay rates. On a resampled month a single start reaches the same optimum as the full start grid, to the last decimal of the log-likelihood.} Table~\ref{tab:blockboot} compares the two. The block standard error of May's clean $n(1\sx)$ is 0.003--0.05, against 0.002--0.003 parametric: about equal for BAC, four to five times larger for most stocks, eight for MDLZ and seventeen for TGT, which holds an earnings day (Section~\ref{sec:activity}). Blocks of one and ten days give the same picture (0.004--0.056 and 0.002--0.025). Across the 112 stock-months the block standard error of $n(1\sx)$ is 0.015 at the median (0.007--0.019 by stock, Table~\ref{tab:masses}) and 0.076 at most, that is, 7.5\% of the estimate at the median and 25\% at most. The difference from the parametric value is genuine day-to-day variation of excitation, which the parametric bootstrap cannot see. For a stock-month's $n(1\sx)$, the block standard error is therefore the one to report, and the parametric one measures estimation noise only. The raw/clean contrast is unaffected. Raw $n(1\sx)$ varies less across resampled months than clean $n(1\sx)$ (standard deviation 0.003--0.024), as it should if the brief-state inflation is stable from day to day. The 95\% block intervals of the raw/clean ratio within one second have lower ends of 1.5 (TGT) and 1.8 (ATVI) and of 2.5 to 7.6 in the six core stocks, and the block medians agree with the point estimates within 0.5.

\begin{table}[!t]
\centering
\caption{Parametric against paired day-block bootstrap, May 2017. Standard deviations of the clean (\Em) $n(1\sx)$ and $n(10\sx)$ across replicates; raw/clean ratio of $n(1\sx)$: point estimate (\Eo{} with $K = 4$), block median and 95\% percentile interval; the same interval for $n(10\sx)$.}
\label{tab:blockboot}
\small
\begin{tabular}{lccccc}
\toprule
& \multicolumn{2}{c}{sd of clean $n(1\sx)$} & sd of clean $n(10\sx)$ & Raw $\div$ clean $n(1\sx)$ & Raw $\div$ clean $n(10\sx)$ \\
\cmidrule(lr){2-3}
Stock & parametric & block & block & point; block [95\%] & block [95\%] \\
\midrule
INTC & 0.003 & 0.011 & 0.012 & 5.4; 5.8 [4.6, 7.0] & 3.1 [2.8, 3.2] \\
MU   & 0.002 & 0.010 & 0.013 & 4.2; 4.5 [4.0, 5.1] & 2.5 [2.3, 2.7] \\
MDLZ & 0.003 & 0.024 & 0.023 & 2.8; 2.9 [2.5, 3.4] & 2.0 [1.9, 2.2] \\
BAC  & 0.003 & 0.003 & 0.005 & 8.0; 8.1 [7.6, 8.7] & 4.5 [4.2, 4.9] \\
WFC  & 0.002 & 0.010 & 0.014 & 4.0; 4.1 [3.7, 4.5] & 2.4 [2.3, 2.7] \\
T    & 0.003 & 0.012 & 0.017 & 4.6; 4.5 [3.8, 5.3] & 2.8 [2.5, 3.2] \\
TGT  & 0.003 & 0.052 & 0.055 & 1.8; 1.7 [1.5, 2.2] & 1.4 [1.3, 1.7] \\
ATVI & 0.003 & 0.015 & 0.012 & 1.9; 1.9 [1.8, 2.0] & 1.5 [1.4, 1.5] \\
\bottomrule
\end{tabular}
\end{table}

\subsection{Excitation across Months, Days and Hours}
\label{sec:activity}

What the clean series shows about price-change clustering is the subject of this subsection. We present the main panel and examine how the clean excitation moves across months, within the day and across days, whether the feed or the outcome itself could produce the relation, and what changes with the tick regime.

\subsubsection{Timescales across stocks and months}
Tables~\ref{tab:main} and~\ref{tab:masses} are the main panel, one fit per stock and month. The specification is \Em{} with $K = 4$, the refractory period equal to $h$ and the full start grid. For BAC it is $K = 3$ and $h = 2\ms$, since BIC supports a fourth component in 5 of its 14 months only. The first table gives the timescales and the second the kernel masses, with the day-block standard errors of Section~\ref{sec:robust}. In the six core stocks, the fitted kernel places its mass at about 10--13 ms (5 ms in MDLZ, 8 ms in BAC), 60--140 ms, 1.5--2.6 s and 27--40 s. The slowest component carries the most excitation (0.21--0.27 of the 0.32--0.56 total) and the fastest 0.05--0.09. TGT and ATVI have the same shape compressed by a factor of two to three, 3--4 ms, 30--40 ms, 0.8--0.9 s and 11--12 s, with the excitation spread evenly across the four scales and a larger total, about 0.65. The kernel mass within one second runs from 0.10 (BAC) to 0.45 (TGT) across stocks, with block standard errors of 0.007--0.019 at the median month, and within ten seconds from 0.16 to 0.58 (Table~\ref{tab:masses}). The representation is stable across months. The interquartile ranges of the memories are narrow. A three-component kernel switches between alternative triples from month to month, because three components fit different subsets of a richer structure in different months, and with four components the switching is gone. The fastest memory of MDLZ, TGT and ATVI is under about 6 ms and marked as unresolved, since Section~\ref{sec:robust} shows that it follows the merge threshold. The remaining components do not.

\begin{table}[!t]
\centering
\caption{Main panel, timescales: \Em{} with $h = 1\ms$ and $K = 4$ ($h = 2\ms$ and $K = 3$ for BAC), one fit per stock and month. Memories: medians over 14 months [interquartile range]. Excitation: median effective excitation per scale, fastest to slowest (Section~\ref{sec:mass}).}
\label{tab:main}
\scriptsize
\setlength{\tabcolsep}{3pt}
\begin{tabular}{lllllc}
\toprule
Stock & Fast & Second & Third & Slow & Excitation per scale \\
\midrule
INTC & 12.7 ms [9.3--15.6] & 136 ms [72--219] & 2.6 s [1.9--3.8] & 39 s [33--46] & 0.05 $\cdot$ 0.04 $\cdot$ 0.10 $\cdot$ 0.25 \\
MU & 12.0 ms [11.3--13.6] & 74 ms [68--80] & 1.9 s [1.4--2.7] & 37 s [29--44] & 0.07 $\cdot$ 0.06 $\cdot$ 0.14 $\cdot$ 0.25 \\
MDLZ & 5.2 ms$^\dagger$ [4.0--8.0] & 61 ms [37--76] & 1.6 s [1.3--1.9] & 29 s [21--36] & 0.09 $\cdot$ 0.11 $\cdot$ 0.15 $\cdot$ 0.21 \\
BAC ($K=3$) & 8.2 ms [7.9--8.5] & 96 ms [83--109] & --- & 27 s [25--32] & 0.06 $\cdot$ 0.03 $\cdot$ 0.22 \\
WFC & 11.5 ms [9.3--13.1] & 76 ms [58--85] & 1.5 s [1.3--1.9] & 28 s [26--32] & 0.07 $\cdot$ 0.04 $\cdot$ 0.11 $\cdot$ 0.27 \\
T & 11.1 ms [10.0--12.5] & 117 ms [88--161] & 2.2 s [1.9--2.8] & 40 s [31--45] & 0.07 $\cdot$ 0.05 $\cdot$ 0.11 $\cdot$ 0.22 \\
TGT & 3.8 ms$^\dagger$ [3.1--4.6] & 39 ms [34--43] & 0.86 s [0.72--0.95] & 11 s [7.3--13] & 0.15 $\cdot$ 0.15 $\cdot$ 0.15 $\cdot$ 0.17 \\
ATVI & 2.7 ms$^\dagger$ [2.3--3.7] & 30 ms [25--37] & 0.81 s [0.74--1.3] & 12 s [10--21] & 0.10 $\cdot$ 0.18 $\cdot$ 0.16 $\cdot$ 0.19 \\
\bottomrule
\multicolumn{6}{l}{$^\dagger$ Under about 6 ms: not resolved; it follows the merge threshold (Section~\ref{sec:robust}).}
\end{tabular}
\end{table}

\begin{table}[!t]
\centering
\caption{Main panel, kernel mass within one and ten seconds and the effective branching ratio, same fits as Table~\ref{tab:main}. $n(1\sx)$ and $n(10\sx)$: median over the 14 months [interquartile range], with the median day-block standard error of a month's estimate in parentheses (Section~\ref{sec:robust}). Eff.\ BR: median [min--max].}
\label{tab:masses}
\small
\begin{tabular}{lccc}
\toprule
Stock & $n(1\sx)$ & $n(10\sx)$ & Eff.\ BR \\
\midrule
INTC & 0.14 [0.11--0.16] (0.017) & 0.24 [0.22--0.29] (0.019) & 0.46 [0.38--0.63] \\
MU & 0.18 [0.15--0.28] (0.014) & 0.31 [0.28--0.43] (0.022) & 0.53 [0.41--0.73] \\
MDLZ & 0.26 [0.23--0.32] (0.019) & 0.40 [0.36--0.47] (0.020) & 0.56 [0.49--0.65] \\
BAC & 0.10 [0.08--0.12] (0.007) & 0.16 [0.14--0.18] (0.011) & 0.32 [0.25--0.49] \\
WFC & 0.18 [0.16--0.20] (0.015) & 0.32 [0.28--0.34] (0.018) & 0.51 [0.41--0.67] \\
T & 0.15 [0.14--0.19] (0.013) & 0.24 [0.24--0.31] (0.020) & 0.40 [0.35--0.68] \\
TGT & 0.45 [0.40--0.48] (0.016) & 0.58 [0.54--0.64] (0.018) & 0.65 [0.53--0.71] \\
ATVI & 0.40 [0.35--0.48] (0.015) & 0.56 [0.47--0.59] (0.016) & 0.64 [0.53--0.68] \\
\bottomrule
\end{tabular}
\end{table}

\subsubsection{Across months}
\begin{figure}[!t]
\centering
\includegraphics[width=\linewidth, alt={Eight small panels, one per stock, each a scatter of kernel mass within one second against clean events per day on log axes. The stock's own fourteen months are coloured with a fitted line and its slope in the title; all 112 stock-months are grey in the background. The within-stock lines rise with activity at similar slopes, and the grey cloud rises at the same slope.}]{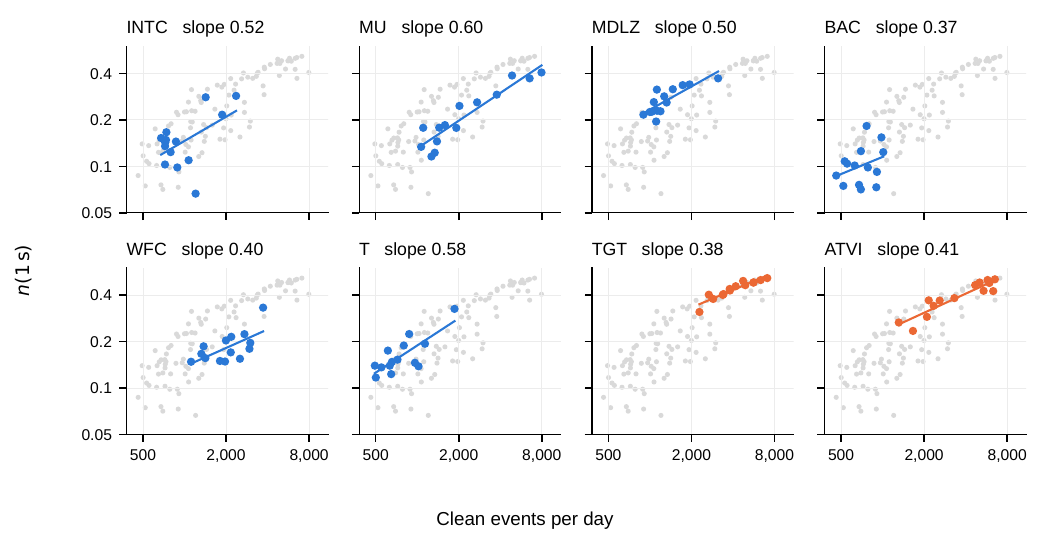}
\caption{Kernel mass within one second against clean events per day, 112 stock-months, one panel per stock (\Em{}, main specification; BAC at $h = 2\ms$ and $K = 3$). Coloured points: the stock's 14 months. Grey points: all 112 stock-months. Line and slope: least-squares fit of $\log n(1\sx)$ on log events per day within the stock. Both axes are logarithmic.}
\label{fig:months}
\alttext{Eight small panels, one per stock, each a scatter of kernel mass within one second against clean events per day on log axes. The stock's own fourteen months are coloured with a fitted line and its slope in the title; all 112 stock-months are grey in the background. The within-stock lines rise with activity at similar slopes, and the grey cloud rises at the same slope.}
\end{figure}

Excitation rises with activity. Within each stock, the month's kernel mass within one second correlates at 0.85 with the log of clean events per day (0.79 for the effective branching ratio), about as strongly as across the eight stocks (0.88--0.91). Busier months are thus more self-exciting in the same way that busier stocks are. The elasticity of $n(1\sx)$ to events per day across months within a stock is 0.50 (0.37--0.60 by stock). Figure~\ref{fig:months} shows the relation stock by stock, with the other stocks' months in grey behind each panel: the eight within-stock slopes are nearly parallel, and the grey cloud has the same slope. The month-to-month variation is real. Within each stock the monthly branching ratio has a standard deviation of 0.05--0.10, six to ten times its parametric bootstrap standard error, and the block bootstrap of Section~\ref{sec:robust} shows that the day-to-day variation behind it is larger still. The timescales move less. Busier months shorten the memories only moderately within a stock (correlations with log activity $-0.11$ to $-0.47$), while the shorter memories of busier stocks are mostly a difference between stocks ($-0.60$ to $-0.87$ across the stock means). Two findings establish that this relation is a property of the data and not of the estimator. The simulation with one kernel and baselines scaled fourfold gives estimates that do not move (Section~\ref{sec:simlimits}), and the raw-event series shows a weaker relation (within-stock correlations 0.68--0.76 against 0.83--0.85 on clean events), as it should if an artifact that does not grow with activity dilutes it. Whether the relation is the feed's, or the outcome measuring itself, is the subject of the checks below.

\subsubsection{Within the day}
The same relation holds within the day. We fit the kernel separately on three windows of every day, 10:00--11:00, 11:00--14:30 and 14:30--15:30, each window treated as its own session with a 60-second burn-in and the fit started from the month's estimates. The first hour has more kernel mass within one second than midday in 91 of the 112 stock-months, with a median ratio of 1.18 (interquartile range 1.06--1.34), and the last hour less, with a median of 0.93 (Table~\ref{tab:byhour}). The timescales are the same in all three windows (median fastest memory 8.3, 8.8 and 8.8 ms; slowest 33, 31 and 26 s), and the weights move, most on the fast components. BAC is the exception in both directions. The first hour is also the busiest, with 1.8--2.4 times midday's events per hour, and the within-day elasticity of $n(1\sx)$ to activity across the three windows is 0.25, about half the across-month value. A month-wide kernel therefore underpredicts the intensity at the open and overpredicts it at the close, which is one of the three sources of the lack of fit examined in Section~\ref{sec:misfit}.

\begin{table}[!t]
\centering
\caption{Kernel by time of day: \Em{} with $K = 4$ and the refractory period (BAC: $K = 3$ and $h = 2\ms$, its main specification), fitted separately on 10:00--11:00 (first hour), 11:00--14:30 (midday) and 14:30--15:30 (last hour) of every day of each stock-month. Columns: ratio of the first hour to midday for $n(1\sx)$ (median over 14 months, min--max), the number of months in which it exceeds 1, the same ratio for $n(10\sx)$, the ratio of the last hour to midday for $n(1\sx)$, and the ratio of events per hour, first hour to midday.}
\label{tab:byhour}
\footnotesize
\setlength{\tabcolsep}{5pt}
\begin{tabular}{lccccc}
\toprule
& \multicolumn{3}{c}{First hour $\div$ midday} & Last $\div$ midday & \\
\cmidrule(lr){2-4}\cmidrule(lr){5-5}
Stock & $n(1\sx)$ & Months $> 1$ & $n(10\sx)$ & $n(1\sx)$ & Activity \\
\midrule
INTC & 1.12 (0.84--1.42) & 11 / 14 & 1.07 & 0.90 & 1.87 \\
MU & 1.23 (1.10--1.60) & 14 / 14 & 1.19 & 0.85 & 2.10 \\
MDLZ & 1.45 (0.59--1.78) & 13 / 14 & 1.28 & 0.90 & 2.36 \\
BAC & 0.84 (0.44--1.10) & 2 / 14 & 0.85 & 1.20 & 1.91 \\
WFC & 1.14 (0.77--1.43) & 10 / 14 & 1.01 & 0.96 & 1.83 \\
T & 1.20 (0.98--1.45) & 13 / 14 & 1.11 & 0.97 & 1.98 \\
TGT & 1.20 (1.06--1.37) & 14 / 14 & 1.12 & 0.96 & 2.13 \\
ATVI & 1.22 (1.09--1.78) & 14 / 14 & 1.16 & 0.86 & 2.08 \\
\bottomrule
\end{tabular}
\end{table}

\subsubsection{Across days}
The relation holds across days as well, in a form that matters for how a month's estimate should be read. To see it without choosing days by their event counts, we do two things. We fit each day on its own with the month's timescales held fixed, so that a day's $n(1\sx)$ is the weight of that day on the month's components. The event-weighted mean of the day values reproduces the month's estimate within 1\% at the median. We then classify days by a calendar fixed before looking at the data: the two sessions after the 2016 election, the nine FOMC decision days, the nine index rebalancing days, and each stock's earnings reaction days, 34 in all. Relative to the median of the ordinary days of the same stock-month, $n(1\sx)$ is 2.03 times higher on the earnings days (interquartile range 1.49--2.36; all 34 above the median), 1.35 times higher on the FOMC days (65 of 72 above) and 1.34 on the election days (14 of 16), and unchanged on the index days (1.04; 40 of 72), whose activity sits in the closing auction, outside our window. The ratios for $n(10\sx)$ are 1.73, 1.31, 1.32 and 1.02. The days that stand out by their counts lead to the same conclusion. In 54 of the 112 stock-months at least one day has more than three times the month's median number of NBBO changes, 84 days in all. Forty of them are calendar days and the rest are news days, December 1, 2017 in six stocks and November 29, 2017 in four, and both groups carry about 1.9 times the ordinary excitation. A month's estimate is therefore partly a few of its days. Without the calendar days, the event-weighted mean of the day values falls by a median 4\% over the 96 stock-months that contain one, by 9\% over the 34 stock-months with an earnings day and by up to 34\% (MDLZ, October 2017), and without the count-flagged days by a median 14\%. Across days within a stock-month, the elasticity of $n(1\sx)$ to the day's event count is 0.34 and to the day's fitted baseline, which the clean process does not generate, 0.27 (0.30 and 0.21 on the ordinary days alone; by stock, 0.30--0.39 for six stocks and 0.08 and 0.10 for INTC and BAC). The simulation's pooling check shows that this is not an artifact of pooling, which is unbiased when the kernel is common to all days. It is also why the model-free standard error of Section~\ref{sec:robust}, which resamples days, is the uncertainty to report for a stock-month's $n(1\sx)$.

\subsubsection{Neither the feed nor the outcome alone}
Two alternatives to reading these gradients as excitation rising with activity have to be excluded. The first alternative is the feed. Under load the SIP's lag could lengthen, more brief states would outlast the threshold and survive the merge, and clean $n(1\sx)$ would rise for that reason. The data say the opposite in every dimension. Refitting the panel with every state shorter than 2 ms merged leaves the across-month relation unchanged (correlation 0.72 against 0.74, elasticity 0.49 against 0.50). The states between 1 and 2 ms add 6\% to $n(1\sx)$, and that excess grows by only 0.017 per doubling of activity. Moreover, the brief states themselves are shorter when the market is busier. Within stocks the median lock duration correlates with activity at $-0.58$ across months and the share of locks longer than a millisecond at $-0.36$. In May 2017 the first hour, with two to three times midday's quotes, has the shortest locks in all eight stocks. Across days within a stock, busier days have shorter locks ($-0.28$) and a smaller tail above the threshold ($-0.25$, negative in every stock). The reason is the one behind the venue lags of Section~\ref{sec:briefstates}: bursts are dominated by the venues with the shortest lag. The lag also fell over the sample, from a median lock of 0.50--0.58 ms before June 2017 to 0.29--0.40 ms after, while activity rose, so the across-month relation could be a trend. The relation is not a trend: the elasticity is 0.46 with a linear month trend, 0.47 with stock and month fixed effects, and 0.41 with the month's lock duration as a regressor, which enters negatively. Finally, in November 2017, the busiest month, the kernel fitted by window of the day is the same on the NBBO rebuilt in exchange time as in SIP time for INTC and MDLZ, and for TGT, with the first-hour excess identical in both clocks (1.32, 1.46 and 1.15). For BAC, whose SIP-time $n(1\sx)$ carries the clock caveat of Section~\ref{sec:robust}, exchange time lowers $n(1\sx)$ least in the first hour, so the intraday load peak is where the SIP-time estimate is inflated least.

The second alternative is the outcome measuring itself. Clean events per day is the rate of the process whose excitation is being estimated, and in a self-exciting model the rate rises with the excitation. Two measures that the stock's own cascade cannot produce give the same direction. The activity of the other seven stocks in the same month correlates positively with a stock's $n(1\sx)$ in every stock (0.27--0.65 by stock), with an elasticity of 0.41. The immigrant rate implied by the fit, $N(1 - n)$ per day, rises with $n(1\sx)$ as well (elasticity 0.54), which a cascade on a fixed immigrant rate would leave unchanged. Own activity dominates when both enter together (0.48 against 0.07), because the stock-specific part of activity carries most of the variation. The elasticity of 0.50 is therefore an upper bound on the exogenous effect, and 0.41 is the elasticity to a measure that the stock's own cascade does not generate. Note that trade counts are not a usable measure across stocks. In May 2017 BAC has three times the trades per day of MDLZ and the lowest excitation, since the number of trades reflects the price level and the tick, not price activity.

\subsubsection{Tick regime}
Table~\ref{tab:regime} sorts the stock-months by the regime of Section~\ref{sec:data}. The brief-state mechanism changes form with the regime but does not go away. In small-tick months locks are rare, 8\% of NBBO changes against 39\% in large-tick months. Yet raw fits still put 44\% of their kernel mass below one millisecond, because the positive-spread brief states of Section~\ref{sec:briefstates} take the place of locks. The raw/clean ratio of $n(1\sx)$ falls from 3.8 to 1.5 across the regimes, because the clean excitation rises (0.16 to 0.48), not because the artifact shrinks (raw $n(1\sx)$ 0.62 to 0.72). Within TGT and ATVI, the two stocks that cross into the small-tick regime, the regime is confounded with activity. Their small-tick months are also their busiest, with two to three times the events per day of their large-tick months (correlation of time at one tick with activity $-0.74$ and $-0.76$), larger clean $n(1\sx)$ and shorter timescales. The lock share and the raw/clean ratio follow the regime directly (correlations $+0.78$ and $+0.94$, $+0.62$ and $+0.77$). Whether the clean excitation does, beyond its relation with activity, these data cannot tell, and we do not claim it.

\begin{table}[!t]
\centering
\caption{Stock-months by tick regime (time at a one-tick spread, excluding locked and crossed states): medians over the stock-months in each class. Raw fits: \Eo{} with $K = 4$; clean: Table~\ref{tab:main}.}
\label{tab:regime}
\small
\begin{tabular}{lccc}
\toprule
& Large-tick (80) & Intermediate (18) & Small-tick (14) \\
\midrule
Time at one tick & 99.99\% & 98.1\% & 66.9\% \\
NBBO changes a day & 2,310 & 5,198 & 10,260 \\
Changes that lock or cross & 39\% & 30\% & 8\% \\
Raw kernel mass below $1\ms$ & 0.46 & 0.47 & 0.44 \\
Raw $n(1\sx)$ & 0.62 & 0.68 & 0.72 \\
Clean $n(1\sx)$ & 0.16 & 0.37 & 0.48 \\
Raw $\div$ clean $n(1\sx)$ & 3.8 & 1.9 & 1.5 \\
Clean fastest / slowest memory & 10.2 ms / 32 s & 3.8 ms / 21 s & 2.7 ms / 9 s \\
\bottomrule
\end{tabular}
\end{table}

\subsection{Goodness of Fit and What the Model Misses}
\label{sec:misfit}

Where the model fitted to the clean series still fails is the subject of this subsection. A Hawkes model describes the clean series much better than the raw one. The better fit says that the clean series is closer to a Hawkes process, not by itself that it is closer to the price. With $K = 3$ for all three definitions, the median KS statistic of the time-rescaled residuals across the panel is 0.017 (up moves) and 0.015 (down) for \Em{}, against 0.025 and 0.028 for \Ei{} and 0.068 and 0.070 for \Eo{} (the main \Em{} fits, with $K = 4$ and the refractory period, give 0.016 and 0.014). The clean model is nonetheless rejected. The bootstrap-calibrated tests of Section~\ref{sec:estimation} reject at the 5\% level in every one of the eight stocks in May 2017, in at least one direction: up moves in six stocks and down moves in five, 11 of 16 tests, with 99 replicates and a smallest attainable p-value of 0.01. In November 2017, the busiest month of the sample, all 16 tests reject: the statistics are 0.014--0.033 for up moves and 0.010--0.027 for down moves, against 95th percentiles of the bootstrap null of 0.004--0.013. The May up-move statistics, 0.015--0.027, lie above the 95th percentile of a correctly specified four-exponential model (0.006--0.013, Section~\ref{sec:simrecovery}) and of a power-law model (0.005--0.016), so the misfit is neither sampling variation nor the kernel's parametric form. Residual diagnostics on the eight May fits, compared with months simulated from the same fits, locate it in three places.

\runin{Over the days, and at the open.} The misfit is spread over the days, not concentrated in a few. The median single-day KS statistic is 0.040--0.082 against 0.020--0.053 in the simulated months, in every stock and both directions. The misfit is largest at the open. The mean residual in the first half hour is 0.83--0.89 for MDLZ, T, TGT and ATVI (0.92--1.03 simulated) and above one in the last half hour. The by-hour fits of Section~\ref{sec:activity} explain this: a month-wide kernel has too little excitation for the first hour and too much for the last. Scoring each hour with the kernel fitted to that hour confirms the attribution. The first-half-hour mean rises to 0.96--0.97 for TGT and ATVI and to 0.91 for MDLZ (0.96--1.00 simulated from the hour's own fit), the last-half-hour mean falls to 0.99--1.02 in the five stocks where it exceeded 1.03, and refitting only the first hour's baseline changes nothing. T's 0.92 is unchanged and sits at the edge of its simulated range. Note that this is not the cold start of a session observed from 10:00 with no history. Scoring the fit from 10:05 with the first five minutes as history leaves the first-half-hour residual mean at 0.88--0.91 for MDLZ, TGT and ATVI, and finer baselines make it worse.

\runin{Bursts shorter than five minutes.} In the busy stocks, events cluster on the scale of tens of seconds more than the fitted model produces. The rate of events 10--100 s after an event, relative to the unconditional rate, is 3.5 for TGT, 2.6 for MDLZ and 2.1 for ATVI and BAC, against 1.5--1.7 in the simulated months (INTC: 1.6 against 1.6). The autocorrelation of the residuals at lags two and three is significant for TGT ($z$ = 4.4 and 6.8), MDLZ, BAC and ATVI. Baselines of 10 and 5 minutes raise the model's figure only to 1.7--1.9, so a finer piecewise-constant baseline does not absorb these bursts, which are shorter than the finest bin. A fifth exponential does not absorb them either: fitted with $K = 5$, the kernel places its new component between 2 ms and 10 s, never at the minute scale, and months simulated from it give the same 10--100 s rate ratio as $K = 4$ (1.5--1.7). These bursts are the minute-scale variation behind the baseline sensitivity of the total in Section~\ref{sec:tradeoff}.

\runin{More reversals than the kernel allows.} The share of events preceded by a move in the opposite direction is higher than in the simulated months in all eight stocks, by 0.01--0.04 overall, by 0.03--0.06 at gaps of 1--10 s and by 0.03--0.09 at 10--100 ms in seven of the eight. The mean residual of a move that follows an opposite move is lower than simulated in all eight. In other words, after a move, the next move in the same direction comes later than the model expects and the next move in the opposite direction sooner. A linear symmetric kernel can raise the cross weight, but it cannot push the intensity of same-direction moves below the baseline plus their self-excitation, and the data show fewer continuations than that. The reversals are the part of the misfit that differs between directions. They give the deficit of residuals around 0.2--0.5 in up moves for INTC and BAC and in down moves for MU and T.

None of the three touches the direction or the existence of the raw/clean contrast, whose mechanism the simulation reproduces. The clean magnitudes, however, are conditional on the linear symmetric model: $n(10\sx)$ through the bursts, which sit where the slowest component lives, and $n(1\sx)$ through the reversals, since a kernel with $a_k \ge 0$ can only load an excess of reversals onto the cross weights. The constraint binds rarely, and where it does not bind it is harmless. Self-excitation sits at zero in 7 of the 112 stock-months (the 1.5--2.6 s component) and cross-excitation in 7 (the slowest component, in the small-tick months of TGT and ATVI). In the other 98 every fitted weight is positive, and then the fit is also the optimum of the same kernel with signed weights and the intensity truncated at zero, the simplest model that allows inhibition. The reason is that at the fitted timescales the log-likelihood is concave in the weights, so an interior optimum is the global one, and truncation can only lower the likelihood where a weight is negative while leaving it unchanged where none is. Allowing inhibition at those timescales would therefore not change $n(1\sx)$ in 98 stock-months, and in the other 14 it could act only through the component at zero, which lies at 1.5 s or beyond. What this argument does not cover is a signed model that prefers other timescales. $n(1\sx)$ is unaffected by the baseline resolution in the 16 stock-months examined (Section~\ref{sec:tradeoff}). Each of the three sources points to a model outside the one used here: a kernel that depends on the time of day, a stochastic or regime-switching baseline for the bursts, and direction-dependent kernel shapes or inhibition for the reversals. We leave them as stated limits of the linear symmetric Hawkes model. Note that the open and the reversals are themselves findings about clean price moves that a raw-event fit, whose residuals are dominated by the brief states, cannot show.

\section{Conclusion}
\label{sec:conclusion}

In this paper, we asked what a Hawkes model fitted to the consolidated quote feed measures, in the stocks where the feed's artifacts are largest. We defined the price-change event as a mid-price level that persists for one millisecond, fitted a multi-exponential Hawkes model with a refractory period of the same length to raw and clean events in eight stocks over fourteen months, simulated price moves from the fitted model through a synthetic consolidated feed, and examined what the clean series shows about price-change clustering.

The sub-millisecond kernel component that raw-event fits find in large-tick stocks, with about half of all events attributed to it, is the feed's construction. It describes how locked and depleted quote states are resolved, and a synthetic feed carrying nothing but the measured brief states reproduces it. The clean definition removes the component. What remains is a kernel whose mass lies between about 10 ms and 30 s, and whose horizon masses and three slower memories are stable across months, across thresholds from one to five milliseconds and across clocks. Raw fits put three times the clean kernel mass within one second, most of it in the first millisecond, while on the total branching ratio the artifact and the dilution by the doubled event count offset, so the total hides a distortion that the horizon masses show. Comparisons of Hawkes estimates from consolidated data across stocks, periods or papers therefore require a stated threshold.

The total branching ratio has a second problem that does not depend on the feed. Its largest part is the slowest kernel component, which trades off against the baseline, so the total moves with the baseline resolution while the kernel mass within one second does not, and within ten seconds only at the finest baselines. The kernel mass within one and ten seconds is what a multi-exponential fit to these data identifies, and we report it with standard errors from a block bootstrap over days.

On the clean series, excitation rises with activity across months, within the day and across days, with market-wide activity as with the stock's own, and not through the feed, whose brief states shorten under load. The timescales move less than the weights. In the two months tested the model is still rejected, in three ways we can locate: stronger excitation at the open than a month-wide kernel allows, bursts shorter than five minutes, and more reversals than a linear symmetric kernel can produce.

What a Hawkes model measures in the consolidated feed therefore depends on the event definition: fitted to raw events it measures the feed, and fitted to events with a stated threshold it measures the clustering of price moves. Two limits set the next steps. First, the sample is eight large-tick stocks over fourteen months of 2017, in a feed whose lag has since fallen and whose tick the 2024 amendments to Rule 612 would change where they apply \citep{sec2024}. The definition applies as it stands to newer consolidated data and to the direct exchange feeds, where the brief states are shorter and the threshold could be lowered, and the first question there is whether the clean kernel of a large-tick stock is the same whichever feed records it. Second, the model is a linear symmetric Hawkes process with a piecewise-constant baseline, and each source of its misfit names a model of clean price moves: a kernel that depends on the time of day for the open, a stochastic or regime-switching baseline for the bursts, and direction-dependent kernels or inhibition for the reversals. The definition makes each of them estimable on consolidated data without first explaining the feed.

\begin{appendices}
\section{The Gated Likelihood}
\label{app:likelihood}

Consider one session $[0, T]$ with events $t_1 < \dots < t_N$ of types $\tau_n \in \{u, d\}$, and the model \eqref{eq:model}--\eqref{eq:kernel} with dead windows $D_n = [t_n, t_n + \delta_n)$, $\delta_n = \min(\delta,\, t_{n+1} - t_n,\, T - t_n)$, so that $D = \bigcup_n D_n$ is a disjoint union and the gate at time $t$ is $D_t = \bigcup_{n:\, t_n < t} D_n$. Write $R_{j,k}(t) = \sum_{t^j_m < t} e^{-\beta_k (t - t^j_m)}$ and $R_{j,k}(t_n^+) = R_{j,k}(t_n) + \mathbf{1}\{\tau_n = j\}$ for the decayed count just after event $n$. The target process has no two events closer than $\delta$, so no event falls in the dead window of an earlier event and the intensity at every event is the ungated one,
\begin{equation}
\lambda_i(t_n) = \mu_{b(t_n)} + \sum_{j} \sum_{k} \alpha_{ij,k}\, \beta_k\, R_{j,k}(t_n),
\label{eq:lam-event}
\end{equation}
where $b(t)$ is the baseline bin containing $t$. The compensator of type $i$ is the integral of the ungated intensity over $[0, T]$ less its integral over the dead windows. The baseline part is $\sum_b \mu_b (|b| - |b \cap D|)$. For the kernel part, the integral of component $k$ of a past type-$j$ event over $[0, T]$ is $1 - e^{-\beta_k (T - t^j_m)}$, and over the dead window $D_n$ the contribution of all type-$j$ events up to and including $t_n$ is
\[
\int_{t_n}^{t_n + \delta_n} \sum_{t^j_m \le t_n} \beta_k e^{-\beta_k (s - t^j_m)}\, ds \;=\; R_{j,k}(t_n^+)\big(1 - e^{-\beta_k \delta_n}\big).
\]
Hence
\begin{equation}
\begin{split}
\int_0^T \lambda_i(s)\, ds = {} & \sum_b \mu_b \big(|b| - |b \cap D|\big) \\
& + \sum_j \sum_k \alpha_{ij,k} \Big[ \sum_{t^j_m} \big(1 - e^{-\beta_k (T - t^j_m)}\big) - \sum_{n=1}^{N} R_{j,k}(t_n^+)\big(1 - e^{-\beta_k \delta_n}\big) \Big],
\end{split}
\label{eq:compensator}
\end{equation}
where the last sum runs over all events of either type. The log-likelihood \eqref{eq:loglik} of the session is the sum over types of $\sum_{n: \tau_n = i} \log \lambda_i(t_n)$ minus \eqref{eq:compensator}, and the month's log-likelihood is the sum over its days, each day starting with $R_{j,k}(0) = 0$.

\runin{Stability.} The gated process needs no stationarity condition. Its intensity is zero on $D_t$ and the windows are disjoint, so at most one event occurs per $\delta$ of time and the rate is bounded by $1/\delta$. Suppose that the gated and the ungated process are both built by thinning the same Poisson random measure on $[0, T] \times [0, \infty)$. Then the monotonicity of the intensity in the history ($\alpha \ge 0$) and the fact that the gate only removes points give the gated process pointwise fewer events than the ungated one. The ungated process has finite moments on a bounded session for any $n$, and is stationary when $n < 1$ with a constant baseline. Nothing in the paper uses stationarity. The cluster interpretation of $n$ is the approximate one of Appendix~\ref{app:mass}.

\runin{Profile likelihood.} For fixed $\boldsymbol\beta$, \eqref{eq:lam-event} is linear in $\boldsymbol\theta = (\mu, a, c)$: with the symmetric kernel, $\lambda_i(t_n) = \mathbf{x}_n^\top \boldsymbol\theta$, where $\mathbf{x}_n$ has a one in the position of $\mu_{b(t_n)}$, the entries $\beta_k R_{i,k}(t_n)$ in the positions of $a_k$ and $\beta_k R_{\bar\imath,k}(t_n)$, for the opposite type $\bar\imath$, in those of $c_k$. The compensator \eqref{eq:compensator} is likewise $\mathbf{c}^\top \boldsymbol\theta$ for a vector $\mathbf{c}$ that collects the bin lengths net of dead time and the bracketed kernel terms, summed over the two types that each kernel entry excites. The log-likelihood $\ell(\boldsymbol\theta \mid \boldsymbol\beta) = \sum_n \log(\mathbf{x}_n^\top \boldsymbol\theta) - \mathbf{c}^\top \boldsymbol\theta$ is concave in $\boldsymbol\theta$, with gradient $\sum_n \mathbf{x}_n / (\mathbf{x}_n^\top \boldsymbol\theta) - \mathbf{c}$, and is maximized over $\boldsymbol\theta \ge 0$ by L-BFGS-B. The design vectors and the decayed counts depend on $\boldsymbol\beta$ only through the per-day recursions, one per component, so each evaluation of the profile likelihood costs $O(KN)$, and the outer Nelder--Mead search over $\log \boldsymbol\beta$ reuses the recursions of recently visited decay rates.

\runin{Residuals.} The time-rescaled residuals of type $i$ are the increments of $\Lambda_i(t) = \int_0^t \lambda_i(s)\,ds$ between consecutive type-$i$ events. $\Lambda_i(t_n)$ is computed from the same quantities: the cumulative baseline up to $t_n$ net of the dead time already elapsed, plus $\sum_j \sum_k \alpha_{ij,k}\big(N_j(t_n^-) - R_{j,k}(t_n)\big)$, less the dead-window terms of \eqref{eq:compensator} for the windows before $t_n$.

\runin{Burn-in.} For the fits by time of day (Section~\ref{sec:activity}) and the diagnostics of Section~\ref{sec:misfit} the likelihood is scored on $[t_b, T]$ only, with the events before $t_b$ kept as history. They enter the recursions, but their own log-intensities are not summed. The baseline integral runs over $[t_b, T]$, the kernel integral of an event $t^j_m < t_b$ runs from $t_b$ (contributing $e^{-\beta_k (t_b - t^j_m)} - e^{-\beta_k (T - t^j_m)}$), and only the dead windows of events at or after $t_b$ are subtracted. The part of the last earlier window that extends past $t_b$, at most $\delta$, is neglected. The main fits use no burn-in.

\runin{Verification.} We test the implementation (Python, with the recursions compiled) against an $O(N^2)$ evaluation of the likelihood from its definition, against numerical integration of the gated intensity for the dead-window and burn-in terms, and against simulation. We generate data by Ogata's thinning method with the dead windows, using the ungated intensity as the dominating rate, which bounds the gated one everywhere, including at the end of each window where the gated intensity jumps back up. Refitted, the data recover the generating parameters, and their residual KS statistics fall where the parametric bootstrap of Section~\ref{sec:estimation} places them under a correct model.

\section{Kernel Mass within a Horizon under the Refractory Period}
\label{app:mass}

Section~\ref{sec:mass} defines $n(H)$ in \eqref{eq:nH} as the kernel mass between an event's own dead window and the horizon, and calls $n(\infty) = \sum_k (a_k + c_k) e^{-\beta_k \delta}$ the effective branching ratio. These are summaries of the fitted kernel. Under the model the number of offspring an event actually produces is smaller, because the gate follows every event: the kernel of an event at $t_m$ is also switched off inside the dead windows of all later events. The number of its direct offspring of either type that the kernel can realize within $H$ on the observed path, with the dead windows of the later events switched off, is
\begin{equation}
\tilde n_m(H) = \sum_{k=1}^{K} (a_k + c_k) \int_{t_m + \delta}^{t_m + H} \beta_k e^{-\beta_k (s - t_m)}\, \mathbf{1}\{s \notin D\}\, ds,
\label{eq:fullygated}
\end{equation}
where $D$ is the union of the dead windows of the events after $t_m$. Averaged over the events of a sample, $\tilde n(H)$ is the fully gated counterpart of $n(H)$, and $n(H) - \tilde n(H)$ is the mass lost to later windows.

The loss is small for a simple reason. For a component with memory $1/\beta_k$ the lost fraction of its mass beyond $\delta$ is the probability, averaged over the kernel's own delay distribution, that a time $s$ after the event falls inside a later dead window, which is about $\delta$ times the conditional rate of further events at lag $s$. That rate is high only at short lags, so the loss concentrates on the fastest component, and even there the dead windows cover a small fraction of the time axis: for a fastest memory of 10 ms and a conditional rate of some tens of events per second at that lag, $\delta = 1\ms$ gives a loss of a few percent of that component's mass, and the slower components lose far less.

Table~\ref{tab:fullygated} evaluates \eqref{eq:fullygated} on the actual event histories of May 2017 for the four simulation stocks, with the main fits. The extra loss is 0.16--2.3\% of the fastest component's mass, and 0.0001--0.005 of the effective branching ratio, at most 1\% of it (TGT, the busiest stock, with the shortest fastest memory). For $n(1\sx)$ and $n(10\sx)$ the loss is of the same size in absolute terms, since it sits almost entirely below one second. The difference is below the parametric standard error of the effective branching ratio (0.005--0.012) in every stock except TGT, where it is about one standard error, and well below the day-block standard errors of Section~\ref{sec:robust}. We therefore report the kernel summaries \eqref{eq:nH} throughout and note here that, under the model, realized excitation is smaller by the amounts in the table.

\begin{table}[!h]
\centering
\caption{Kernel summaries against the fully gated quantities \eqref{eq:fullygated}, evaluated on the May 2017 event histories with the main fits.}
\label{tab:fullygated}
\small
\begin{tabular}{lccc}
\toprule
Stock & Extra loss of the fastest component & Effective BR, \eqref{eq:nH} & Fully gated, \eqref{eq:fullygated} \\
\midrule
INTC & 0.16\% & 0.4372 & 0.4371 \\
BAC  & 0.49\% & 0.3546 & 0.3543 \\
MDLZ & 0.72\% & 0.4866 & 0.4857 \\
TGT  & 2.3\%  & 0.5936 & 0.5882 \\
\bottomrule
\end{tabular}
\end{table}

One further point concerns the statement in Section~\ref{sec:definition} that the branching ratio equals the expected fraction of events that are offspring, not immigrants. Without the gate it holds exactly for a stationary process, since with the symmetric kernel every event produces $\sum_k (a_k + c_k)$ expected offspring whichever its type. With the gate it holds for the fully gated quantity, and approximately for the kernel summary. The simulation of Section~\ref{sec:simrecovery} bounds the approximation: the number of events the fitted model attributes to the baseline exceeds the true number of immigrants by 2--14\%, almost all of it from the slow-component bias and little from the gate.

\section{Price Units at Three Horizons}
\label{app:priceunits}

Table~\ref{tab:priceunits} gives the values drawn in Figure~\ref{fig:priceunits} at 100 ms, 1 s and 10 s, with the standard errors of the observed means, which Section~\ref{sec:rawclean} quotes.

\begin{table}[!h]
\centering
\caption{Signed change of the mid-price after an up move, in ticks, May 2017: observed in the data (standard error from the daily means) and implied by the fitted model, computed as the same statistic on five months simulated from it. Raw: \Eo{} with $K = 4$; clean: \Em{}, main specification.}
\label{tab:priceunits}
\scriptsize
\setlength{\tabcolsep}{3.5pt}
\begin{tabular}{llcccccc}
\toprule
& & \multicolumn{2}{c}{Within 100 ms} & \multicolumn{2}{c}{Within 1 s} & \multicolumn{2}{c}{Within 10 s} \\
\cmidrule(lr){3-4}\cmidrule(lr){5-6}\cmidrule(lr){7-8}
Stock & & Data (SE) & Model & Data (SE) & Model & Data (SE) & Model \\
\midrule
INTC & \Eo & $+0.205$ (0.003) & $+0.589$ & $+0.188$ (0.006) & $+0.576$ & $+0.117$ (0.011) & $+0.476$ \\
     & \Em & $-0.001$ (0.003) & $+0.011$ & $-0.019$ (0.007) & $-0.007$ & $-0.091$ (0.014) & $-0.059$ \\
BAC  & \Eo & $+0.146$ (0.004) & $+0.392$ & $+0.144$ (0.006) & $+0.386$ & $+0.090$ (0.014) & $+0.338$ \\
     & \Em & $+0.011$ (0.002) & $+0.029$ & $+0.006$ (0.005) & $+0.023$ & $-0.060$ (0.015) & $-0.019$ \\
MDLZ & \Eo & $+0.170$ (0.009) & $+0.464$ & $+0.171$ (0.014) & $+0.452$ & $+0.152$ (0.020) & $+0.408$ \\
     & \Em & $-0.014$ (0.009) & $-0.020$ & $-0.016$ (0.014) & $-0.037$ & $-0.036$ (0.022) & $-0.044$ \\
TGT  & \Eo & $+0.108$ (0.008) & $+0.287$ & $+0.102$ (0.012) & $+0.289$ & $+0.129$ (0.019) & $+0.321$ \\
     & \Em & $-0.052$ (0.009) & $-0.051$ & $-0.063$ (0.014) & $-0.053$ & $-0.035$ (0.021) & $-0.025$ \\
\bottomrule
\end{tabular}
\end{table}

\end{appendices}

\section*{Data availability statement}
The quote and trade data underlying this article were provided by the New York Stock Exchange (Trade and Quote database) through Wharton Research Data Services under licence and cannot be shared publicly. The query scripts that extract them, the code that builds the event series and fits the models, and the derived data on which the tables and figures rest (fitted parameters for every stock-month, bootstrap and simulation outputs) are available in the \texttt{hawkes-nbbo-events} repository (\url{https://github.com/hyoeun-lee/hawkes-nbbo-events}) and archived at Zenodo (\url{https://doi.org/10.5281/zenodo.23675006}).

\section*{Acknowledgements}
This work made use of the Illinois Campus Cluster, a computing resource that is operated by the Illinois Campus Cluster Program (ICCP) in conjunction with the National Center for Supercomputing Applications (NCSA) and which is supported by funds from the University of Illinois Urbana-Champaign.

\bibliographystyle{oup-abbrvnat}
\bibliography{references}

\end{document}